\documentclass[a4paper,11pt]{article}
\pdfoutput=1 

\usepackage{jinstpub} 
\let\ifpdf\relax
\usepackage{color}
\usepackage[normalem]{ulem}
\usepackage{lineno}
\usepackage{float}
\usepackage{graphicx}
\usepackage{subcaption}
\usepackage{mwe}
\usepackage{amsmath}
\usepackage{amssymb}
\usepackage{siunitx}
\usepackage{afterpage}
\usepackage{newtxmath,newtxtext}

\usepackage{helvet}
\def\Put(#1,#2)#3{\leavevmode\makebox(0,0){\put(#1,#2){#3}}}

\newlength{\figwidth}
\newlength{\fighalfwidth}
\title{\boldmath \center \LARGE
The ProtoDUNE-SP LArTPC Electronics Production, Commissioning, and Performance
}

\author[b]{D.~Adams}
\author[b]{M.~Bass}
\author[b]{M.~Bishai}
\author[m]{C.~Bromberg}
\author[m]{J.~Calcutt}
\author[b]{H.~Chen}
\author[b]{J.~Fried}
\author[f]{I.~Furic}
\author[b]{S.~Gao}
\author[a]{D.~Gastler}
\author[l]{J.~Hugon}
\author[b]{J.~Joshi}
\author[b]{B.~Kirby}
\author[s]{F.~Liu}
\author[m]{K.~Mahn}
\author[d]{M.~Mooney}
\author[h]{C.~Morris}
\author[c]{C.~Pereyra}
\author[e]{X.~Pons}
\author[b]{V.~Radeka}
\author[b]{E.~Raguzin}
\author[m]{D.~Shooltz}
\author[b]{M.~Spanu}
\author[b]{A.~Timilsina}
\author[v, 1]{S.~Tufanli}
\author[l]{M.~Tzanov}
\author[b]{B.~Viren}
\author[b]{W.~Gu}
\author[p]{Z.~Williams}
\author[n]{K.~Wood}
\author[b]{E.~Worcester}
\author[b]{M.~Worcester}
\author[n]{G.~Yang}
\author[b]{J.~Zhang}

\affiliation[a]{Boston University, Boston, MA, 02215, USA}
\affiliation[b]{Brookhaven National Laboratory (BNL), Upton, NY, 11973, USA}
\affiliation[c]{University of California (Davis), Davis, CA 95616, USA}
\affiliation[d]{Colorado State University, Fort Collins, CO 80523, USA}
\affiliation[e]{European Organization for Nuclear Research (CERN), Geneva, 1211, Switzerland}
\affiliation[f]{University of Florida, Gainesville, FL 32611, USA}
\affiliation[h]{University of Houston, Houston, TX 77204, USA}
\affiliation[l]{Louisiana State University, Baton Rouge, LA 70803, USA}
\affiliation[m]{Michigan State University, East Lansing, MI 48824, USA}
\affiliation[n]{Stony Brook University, Stony Brook, New York 11794, USA}
\affiliation[p]{University of Texas (Arlington), Arlington, TX 76019, USA}
\affiliation[s]{Tsinghua University, Beijing, 10084, China}
\affiliation[v]{Yale University, New Haven, CT 06520, USA}
\affiliation[1]{Now at European Organization for Nuclear Research (CERN), Geneva, 1211, Switzerland}

\emailAdd{mspanu@bnl.gov}

\abstract{
The ProtoDUNE-SP detector is a large-scale prototype of the Single-Phase (SP) Liquid 
Argon Time Projection Chamber (LArTPC) design proposed for the Deep Underground Neutrino 
Experiment (DUNE). 15,360 LArTPC wires are instrumented with low electronic noise pre-amplifier 
and digitization ASICs integrated into Front End Motherboards (FEMBs) operating at cryogenic 
temperature within the cryostat. The large number of electronics channels and high performance 
specifications required a large-scale production electronics quality control effort, careful 
installation into Anode Plane Assemblies (APAs), and rigorous detector commissioning. This 
successful collaboration-wide effort achieved a working LArTPC electronics channel percentage 
of 99.7\% (15,318 of 15,360 channels in total), whose operating performance exceeded expectations. We summarize the ProtoDUNE-SP cold electronics design and quality control, 
installation, and commissioning efforts that enabled this excellent electronics performance. 
}

\keywords{ProtoDUNE, Electronics}

\begin{document}
\maketitle
\flushbottom

\pagebreak

\section{Introduction} \label{sec:introduction}

The Deep Underground Neutrino Experiment (DUNE) is a next-generation neutrino oscillation experiment. DUNE's scientific goals include precise measurements of the parameters governing  neutrino oscillation, in addition to sensitivity to CP-violation discovery as well as sensitivity to neutrinos from core-collapse supernovae and a variety of BSM physics including baryon number violating processes.
DUNE will consist of an intense long-baseline neutrino beam from Fermi National Accelerator Laboratory in Batavia, Illinois, to the Sanford Underground Research Laboratory in South Dakota, approximately 1300 kilometers downstream of the source~\cite{dune-idr-v1}. A Near Detector (ND) installed at Fermilab will record particle interactions near the beam source and a massive Far Detector (FD) consisting of four Liquid Argon Time Projection Chambers (LArTPC) holding in total around 68~ktons LAr, will be constructed at the Sanford Lab site. 
The collaboration has undertaken an extensive prototype program (protoDUNE) at the CERN Neutrino Platform facility to establish the design and performance of two variants of the LArTPC technology: Single-Phase~\cite{dune-idr-v2} and Dual-Phase~\cite{dune-idr-v3}.

\begin{figure}[h!]
\centering
\includegraphics[scale=0.17]{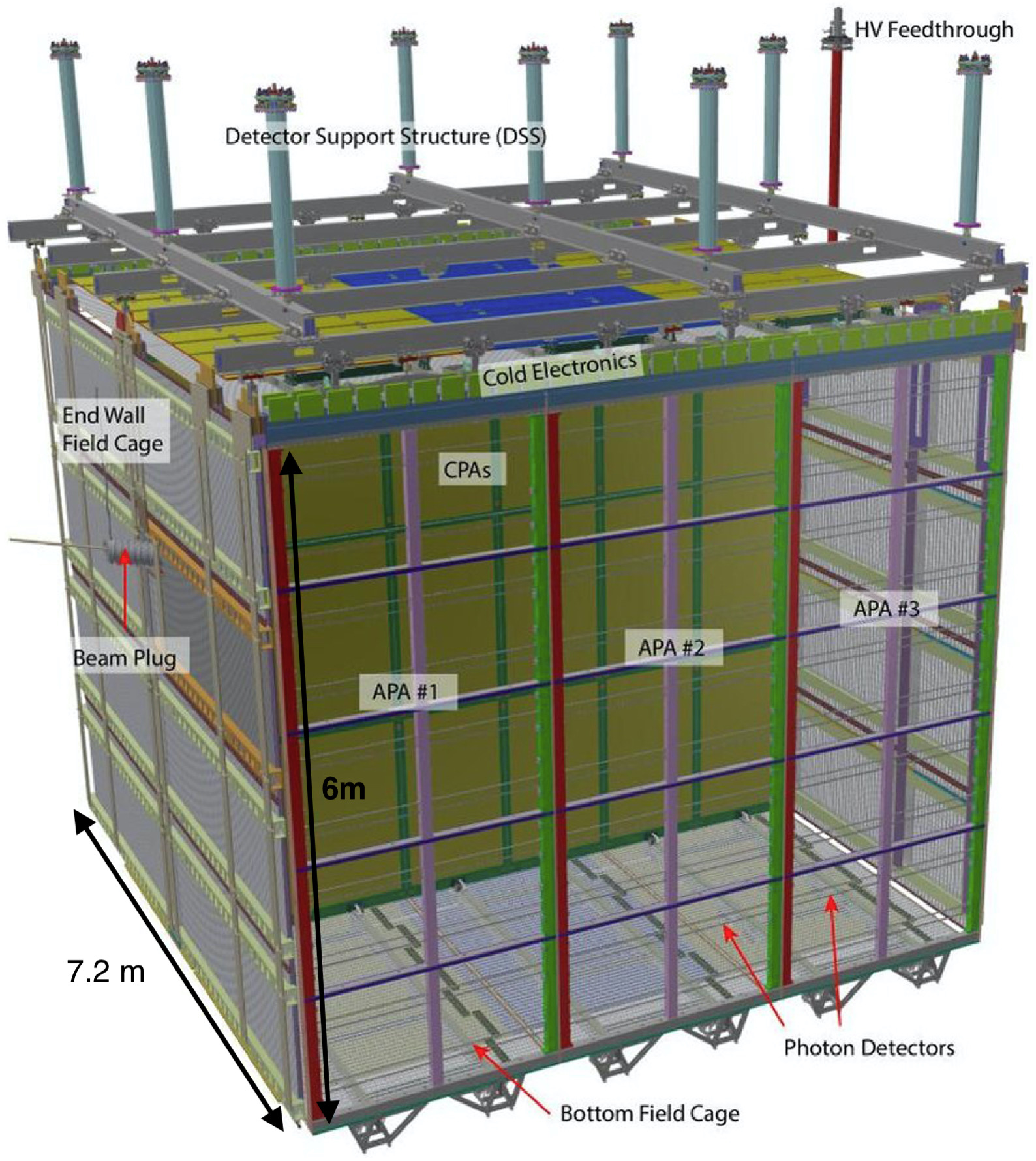}
\caption{ProtoDUNE Single-Phase LArTPC schematic design. The TPC readout Cold Electronics are
indicated by the green boxes along the top of each APA.}
\label{fig:intro-protodune}
\end{figure}

The aim of the protoDUNE program is to better define the production and installation procedures for the DUNE FD as well as accumulate test-beam data at CERN in order to measure the response of the detector to 
different particles at energies in the $0.5-7$~GeV range.
Containing 770~tons of LAr (411~tons active volume) protoDUNE Single-Phase (SP) consists of six full-size 
Anode Plane Assemblies (APAs) for a total of 15,360 TPC sense wire and electronics channels~\cite{protodune-sp-tdr}. 
The 500~V/cm electric field is produced by three Cathode Plane Assemblies (CPA) installed 
in the inner part of the TPC for a total of 2$\times$3.6~meter drift regions, each observed by
three APAs, as shown in Figure~\ref{fig:intro-protodune}. 
Sixteen field cage aluminium profiles maintain a uniform electric field between the cathode and anode. 
The top and bottom of the TPC are equipped with perforated stainless steel ground planes 
to ensure no field outside the active volume.
The detector is located in an extension to the EHN1 hall at CERN and it took its first beam-data from the new H4-VLE beam line before the LHC long shutdown at the end of 2018. The H4-VLE beam line comes from an extension of the secondary $+80$~GeV/c pions beam line, which comes in turn from a first extension of the $400$~GeV/c primary beam from SPS. It consists of tertiary $e^{-}$, p, ~$\mu^{+}$, $\pi^{+}$ beam with momentum range from a $0.5$ to $7$~GeV/c.\\
Electronics noise is characterized by equivalent noise charge (ENC) which is defined as the number of electrons collected at the input of the readout amplification required to produce a signal of magnitude equal to a measured noise RMS. \\
For the entire drift region to be fully active the ENC is required to be less than $\sim1/9$ that of a signal arising from a minimum ionizing particle.  This corresponds to an ENC < 1000 $e^-$ in the case that LAr has a purity such that drifting electrons have a mean lifetime of 3 msec. \\ 
To achieve this, a TPC readout "Cold Electronics" design integrated with the detector electrodes 
has been developed for cryogenic temperatures ($77-89$~K). In this configuration the length of signal carrying wires may be minimized and thermal noise is reduced, so that the ENC is independent of the fiducial volume and lower than with readout electronics at room temperature~\cite{Radeka:2011zz}.

\subsection{TPC Readout System Design}\label{sec:introduction:protodune_elec}

\begin{figure}[h!]
\centering
\includegraphics[scale=0.6]{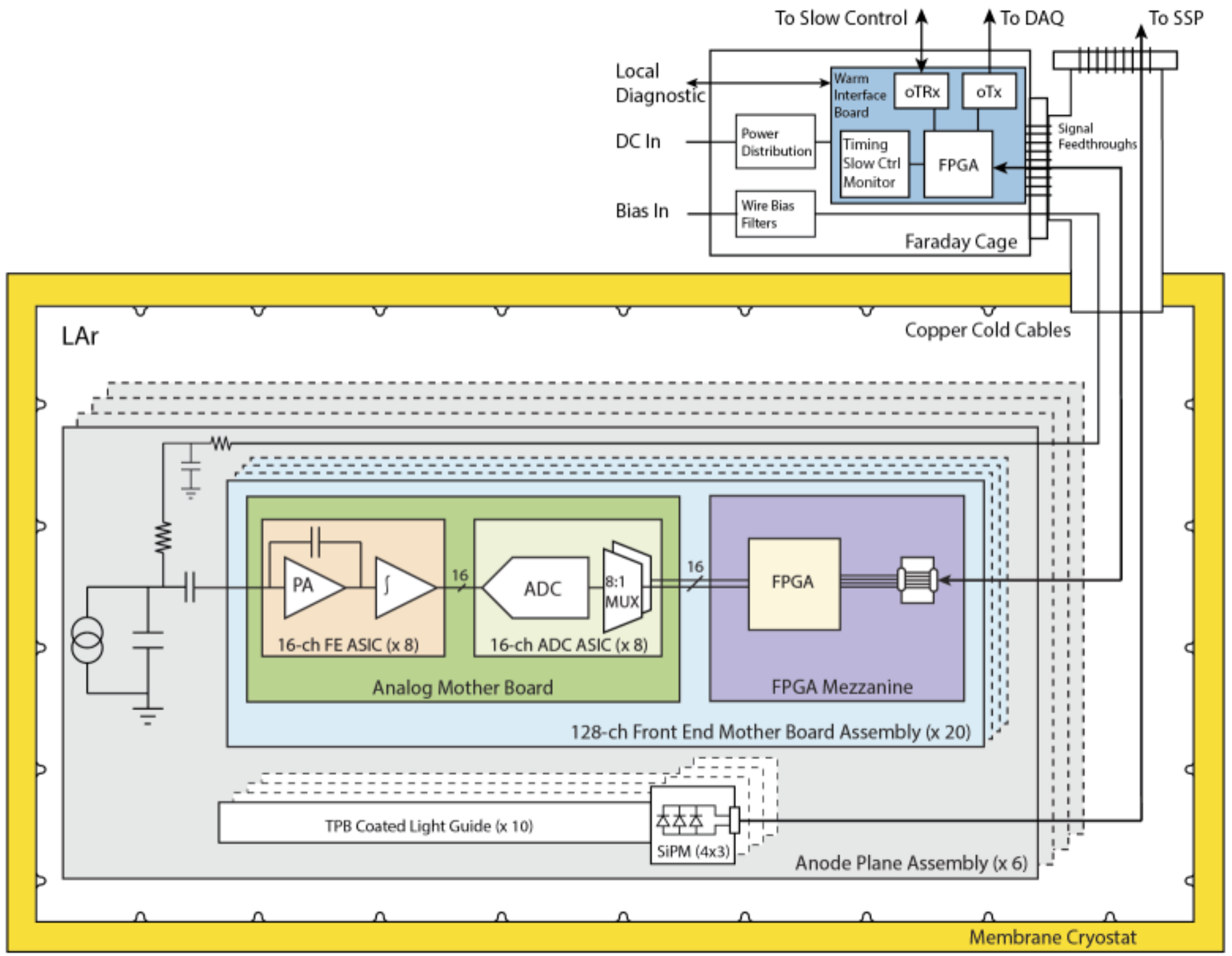}
\caption{The ProtoDUNE-SP CE architecture. All components inside the Membrane Cryostat
are operating submerged in LAr at 89K.}
\label{fig:ce-scheme}
\end{figure}

The ProtoDUNE-SP Cold Electronics (CE) system is shown in 
Figure~\ref{fig:ce-scheme}~\cite{protodune-ce1}~\cite{protodune-ce2}.
The CE provides deadtimeless signal handling and transmission from the detector electrodes
directly on the Anode Plane Assembly (APA), represented in Figure \ref{fig:ce-scheme} by the resistor/capacitor network outside the FEMB, until the data is received by the ProtoDUNE-SP Data Aquisition (DAQ) and Slow Control computers over optical fibers, represented in Figure \ref{fig:ce-scheme} by the arrows from the Warm Interface Board outside of the cryostat.

\subsubsection{Cold Electronics}
\label{ce}

The cryogenic elements of the CE system consist of 20 Front-End Mother Boards (FEMBs) 
installed close to the wire electrodes on top of each APA, which are composed of three wire
planes (2 induction and one collection plane) for a total of 15,360 wires. 
The FEMB amplify, shape, digitize, and transmit all the TPC wires signals to the warm interface electronics through cold data cables. Each 
FEMB contains one Analog Motherboard, which is assembled with eight 16-channel analog 
Front-End (FE) ASICs~\cite{fe-asic}, which provide amplification and pulse shaping,
and eight 16-channel Analog to Digital Converter (ADC) ASICs for a total of 128 
channels readout per FEMB, as listed in Table~\ref{tab:ce-elements}. Both the FE and
ADC ASICs are custom circuits designed at Brookhaven National Laboratory (BNL) 
implemented with the TSMC 180~nm CMOS process and operate at 1.8~Volts,
with very low power consumption to increase the ASIC lifetime in cold,
in order to operate the DUNE FD without significant loss of channels for the 
20+ years required by the physics program~\cite{Shaorui:cmos}. \\
Because the FE ASIC amplifier inputs are attached directly to the APA, ENC from
additional capacitance is minimized. Additionally, due to the CMOS static 
characteristic at cryogenic temperature~\cite{Shaorui:cmos}, the ENC of these ASICs decreases 
at cryogenic temperature, enabling very low ENC operation. Finally, because the
signals are digitized before transmission outside the cryostat, the cryostat
penetrations for signal feed-through (a possible source of excess noise) are 
simplified and the design of the CE system is uncoupled from elements of 
the TPC design, \textit{e.g.} cable length from APA to signal feed-through.

\begin{table}
	\centering
\begin{tabular}[c]{l | l | r}
	\centering
Element                                   &  Quantity            &  Channels per element   \\
\hline
Front-End Mother Board (FEMB)             & 120, 20 per APA      & 128     \\
FE ASIC                                   & 960, 8 per FEMB      & 16          \\
ADC ASIC                                  & 960, 8 per FEMB      & 16          \\
FPGA                                      & 120, 1 per FEMB      & 128          \\
Cold cables                               & 120, 1 per FEMB      & 128      \\
\hline
CE feedthrough                            & 6, 1 per APA         & 2560         \\
Signal flange                             & 6, 1 per APA         & 2560      \\
Warm Interface Electronics Crate (WIEC)   & 6, 1 per APA         & 2560         \\
Warm Interface Board (WIB)                & 30, 5 per WIEC       & 512        \\
Power and Timing Card (PTC)               & 6, 1 per WIEC        & 2560         \\
Power and Timing Backplane (PTB)          & 6, 1 per WIEC        & 2560 \\
\end{tabular}
\caption{Components of the CE system.}
\label{tab:ce-elements}
\end{table}

\begin{figure}[h!]
\centering
\includegraphics[scale=0.6]{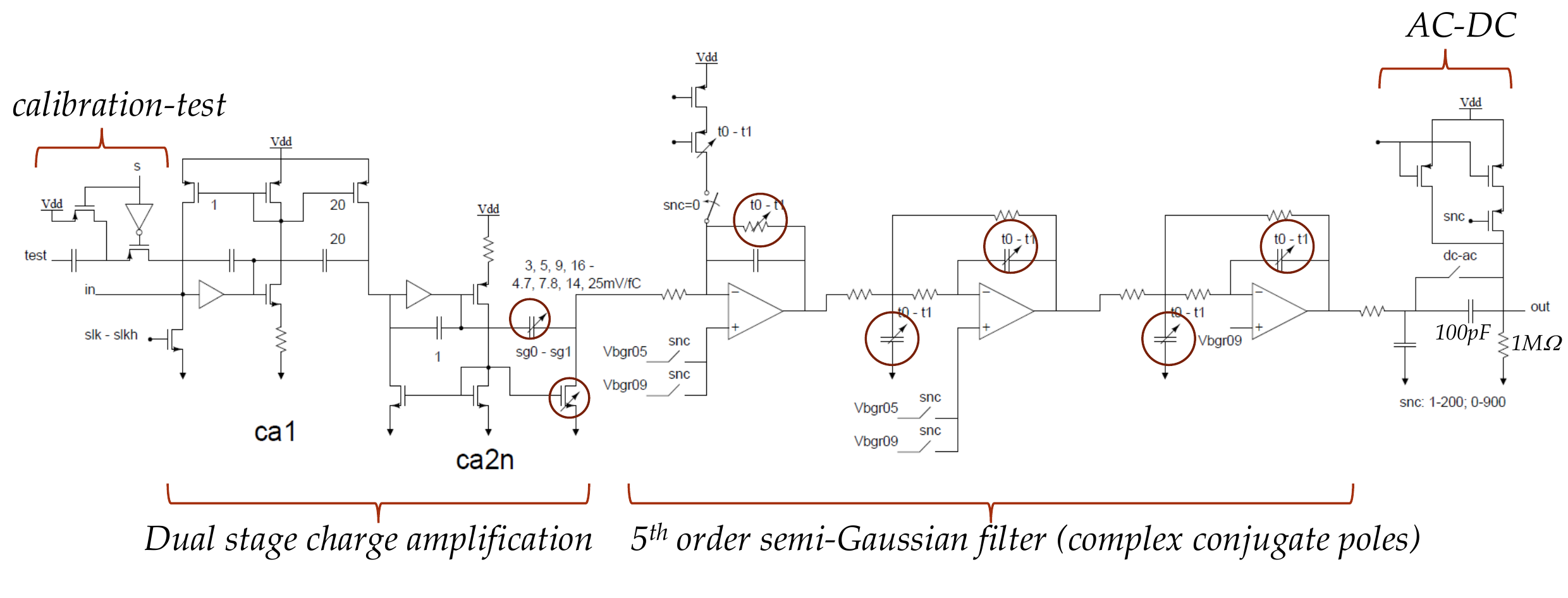}
\caption{Channel schematic of the FE ASIC, which includes a dual-stage charge amplifier 
and a 5th-order shaper with complex conjugate poles. Circuits in red circles are 
programmable to allow different gain and peaking time settings.}
\label{fig:fe-asic-circuit}
\end{figure}

The FE ASIC is three design revisions from the version of the ASIC deployed in the 
MicroBooNE detector~\cite{noise_filter_paper}.
Each FE ASIC channel has a dual-stage charge amplifier circuit with a programmable gain 
selectable from 4.7, 7.8, 14 and 25~mV/fC (full scale charge of 55, 100, 180 and 300~fC),
a 5th-order anti-aliasing shaper with programmable time constant (peaking time 
0.5, 1, 2, and 3~$\mathrm{\mu}$s), an option to enable AC coupling, and a baseline 
adjustment for operation at either 200~mV for the unipolar pulses on the collection 
wires or 900~mV for the bipolar pulses on the induction wires, as shown in 
Figure~\ref{fig:fe-asic-circuit}. Each FE ASIC also has an adjustable pre-amplifier 
leakage current selectable from 100, 500, 1000, and 5000~pA. The leakage current is necessary to keep the feedback loop of the amplifier working properly with an adaptive continuous reset. The adjustment is necessary to accommodate the current caused by wire motion with bias voltages applied on wire planes in the TPC. The estimated power 
dissipation of a FE ASIC is about 5.5~mW per channel at 1.8~V.

Each FE ASIC contains a programmable pulse generator with a 6-bit DAC for electronics 
calibration, which is connected to each channel individually via an injection capacitor
labelled \textit{calibration-test} in Figure~\ref{fig:fe-asic-circuit}.
The injection capacitor is 184~fF at room temperature and 183~fF at 77K, with measured
channel-to-channel variation typically less than 1\%. At cryogenic temperature the packaging of the FE ASIC puts excessive pressure on the ASIC chip. The version of the FE ASIC used in ProtoDUNE-SP responds 
to this stress with a channel dependent non-uniform lowering, called "drop-out", by up to 150 mV of its collection 
mode baseline of 200mV. This results 
in some channels on a few ASICs operating below the minimum acceptable voltage in LAr, which 
compromised the channel performance. In addition, when the baseline is that low, the amplifier is no longer in its linear range. This issue was not observable at room temperature. \\
The ADC ASIC has 16 independent 12-bit digitizers performing at speeds up to 2 
megasamples per second (MS/s), local buffering, and an 8:1 multiplexing stage with 
two pairs of serial readout lines in parallel. Each ADC samples the input voltage in 
the range $0.2-1.6$~V and passes the digitized sample to a built-in FIFO block, 
32 bits deep and 192 (16$\times$12) bits wide, and has full and empty indicator flags,
needed for interfacing to the FPGA. The estimated power dissipation of an ADC ASIC is 
less than 5~mW per channel at 1.8~V. 

\begin{figure}[h!]
\centering
\includegraphics[scale=0.6]{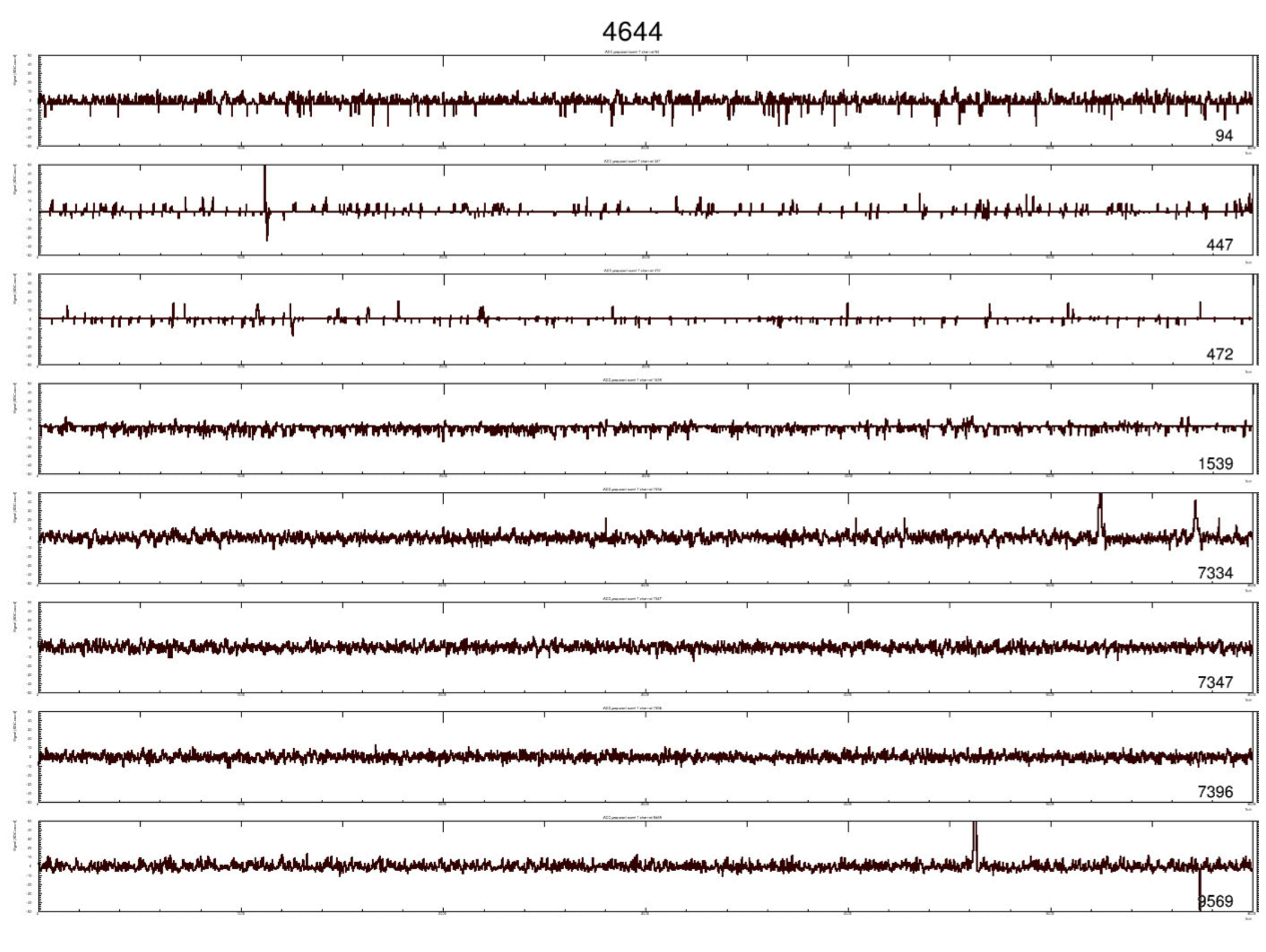}
\caption{Representative waveforms from ProtoDUNE-SP run 4644, with the y-axis from 
-50 to 50 ADC counts and the x-axis in time-ordered 500~ns ticks. The top four show sticky
code behavior from the ADC near the channel baseline, with channels 447 and 472 severely
impacted. For comparison, the bottom four channels do not exhibit sticky code effects, 
showing the normal level of RMS fluctuation near the baseline.}
\label{fig:adc-asic-sc}
\end{figure}

 The version of the ADC ASIC used in ProtoDUNE-SP suffered from several issues at
cryogenic temperature, which were non-observable at room temperature. The most
significant issue with this ASIC was "sticky codes,'' in which certain ADC values
would be preferentially populated by the ADC independent of the input voltage, 
causing the readout channel to appear to ``stick'' at a particular value, 
as shown in Figure~\ref{fig:adc-asic-sc}.
The "domino" architecture~\cite{domino-adc} used in this design relies on excellent
transistor matching, which is difficult to simulate at LAr temperature. The flaw in 
this ADC design was identified as a failure of transistor matching at the transition 
from digitizing the 6 most significant bits to the 6 least significant bits: 
therefore, the sticky codes tended to prefer multiples of 0 and 63 in the ADC 
dynamic range \footnote{A subsequent revision of the FE ASIC was completed after the ProtoDUNE-SP production run that successfully removed the baseline drop-out due to packaging stress. Due to the multiple issues in the domino design ADC ASIC it was not further revised, and a new pipeline design ADC ASIC is now in development by the DUNE collaboration, with a first prototype run having been completed \cite{dune-tdr}. }.\\
A commercial Altera Cyclone IV FPGA, assembled on a mezzanine card which was attached to 
the Analog Motherboard, provided clock and control signals to the FE and ADC ASICs. 
The Cyclone IV was not designed for operation at cryogenic temperature; its 
performance in $77-89$K was validated by standalone tests in liquid nitrogen (LN2).
The FPGA also further serialized the 16 data streams from the ADCs into four 1.25~Gbps 
links for transmission to the warm interface electronics (Section~\ref{wie}).
The FPGA could also provide a calibration pulse to each FE ASIC channel via the
same injection capacitor as used for the internal FE ASIC DAC, as a cross-check
for the electronics calibration. In order to program the FPGA upon start-up, 
the FPGA mezzanine card contains a commercial Altera EPCS64SI16N 64~Mbit 
flash memory chip, which was also not designed for cryogenic operation and required standalone validation tests in LN2.\\
Two sets of 7-meter long cold cable bundles provide power, clock, control, and data 
signals between the CE flange and the FEMBs.
The cold data cables for each FEMB contain 12 twin-axial 26~AWG 
seperately-shielded cables carrying the following differential signals (shown in 
Figure~\ref{fig:pdune-wib}): four 1.25~Gbps data links from the FEMB to the WIB, four 
JTAG programming signals as a backup to program the FEMB FPGA, and two system clocks at
100~MHz and 2~MHz and two I$^2$C-like control links from the WIB to the FEMBs.
The LV power cables contained 9 twisted-pair 20 AWG wires, carrying 1.5V, 2.5V, 
3.0V, 4.2V and 5V. These voltages were futher stepped down by linear power regulators
onboard the FEMB. Shorting caps for both types of cable were designed to make a
low-impedance connection between the signal lines and ground return lines to prevent
charge accumulation on the cables during handling, which could cause electrostatic discharge (ESD) damage to the FEMBs. 

\subsubsection{Warm Interface Electronics}
\label{wie}

\begin{figure}[h!]
\centering
\includegraphics[scale=1.4]{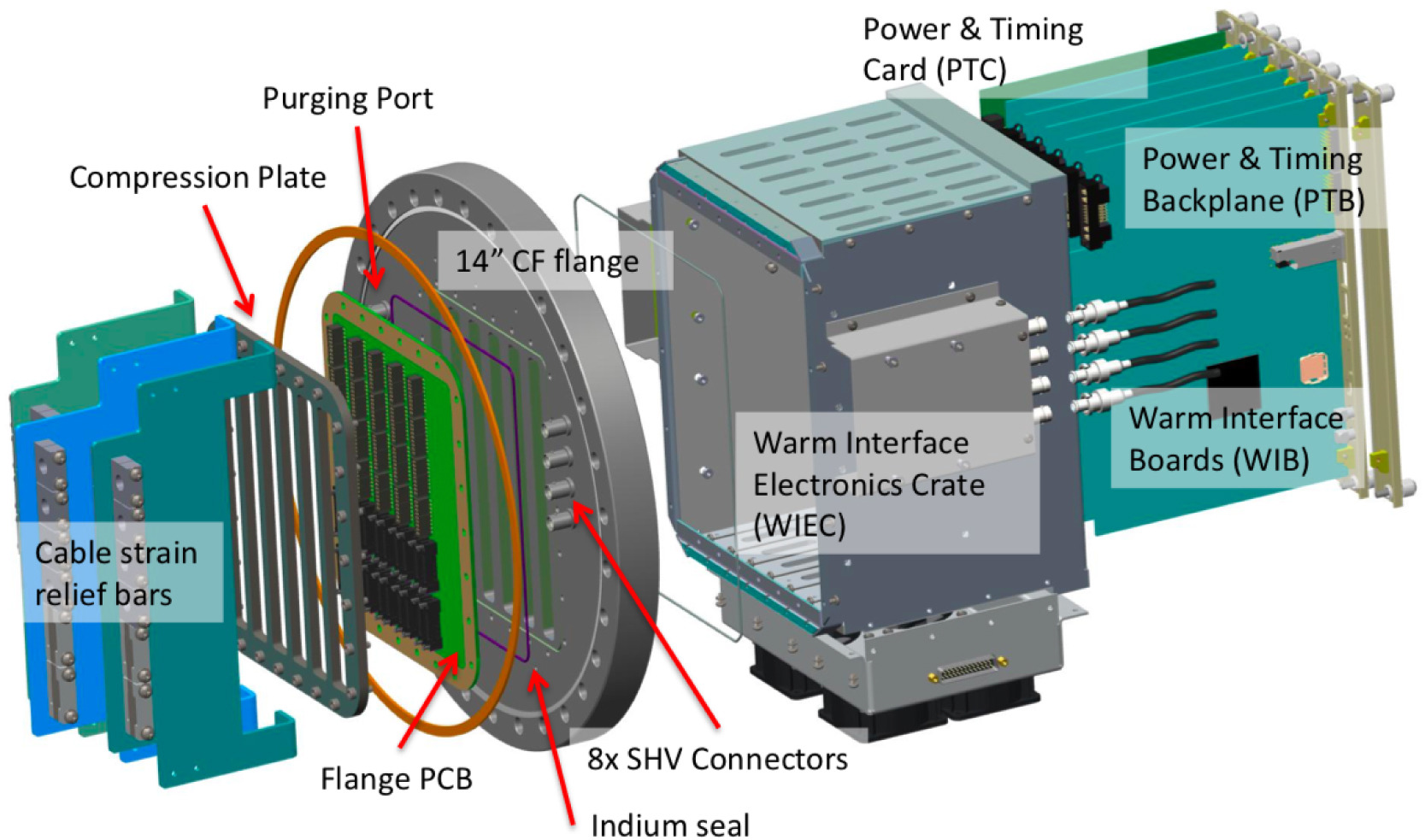}
\caption{Schematic design of the WIEC and components of the warm interface 
electronics.}
\label{fig:ce-flange}
\end{figure}

\begin{figure}[h!]
\centering
\includegraphics[scale=0.4]{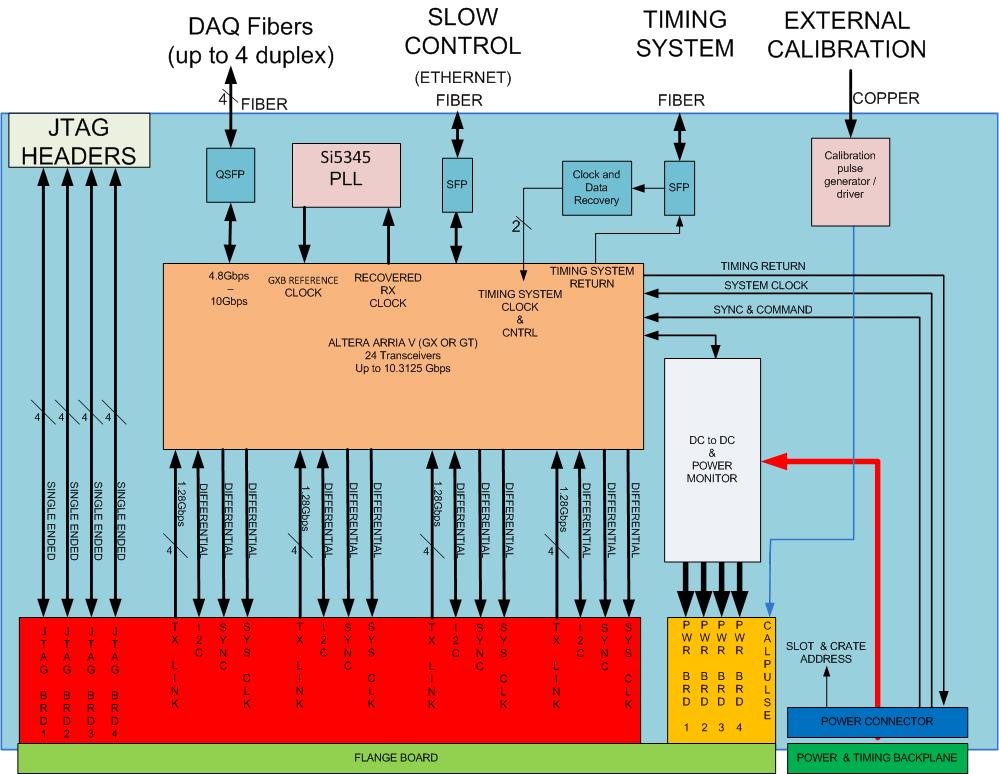}
\caption{Schematic design of the WIB. Front panel inputs are at the top of the image.
12~V power and timing signals are carried on the PTB. Signals to and from up to four
FEMBs are carried on the flange PCB.}
\label{fig:pdune-wib}
\end{figure}

 The warm interface electronics is the interface between the CE and DAQ/timing systems. 
It is housed in the Warm Interface Electronics Crates (WIECs) attached directly to 
the signal flanges. Each WIEC has one Power and Timing Card (PTC), five Warm Interface 
Boards (WIBs) and a passive Power and Timing Backplane (PTB), as shown in 
Figure~\ref{fig:ce-flange}. A flange PCB board with only surface mount components to 
prevent air leakage into the cryostat carries all CE signals in and out of the 
cryostat. \\
The PTC provides both a bidirectional fiber optical link to the timing system and the
input of the 48~V power to the WIEC. It steps down the 48~V to up to six 12~V lines 
and fans out 12~V power and clock signals to WIBs over the PTB. The WIB provides both
local power and control for up to 4 FEMBs, and includes a real-time digital 
diagnostic readout on a dedicated gigabit Ethernet (GbE) UDP link. Each WIB is controlled 
by an Altera Arria V GT FPGA and is powered by 12~V. DC/DC converters onboard the WIB 
further step down the 12~V to 9 power lines between 1.5 and 5~V for each 
FEMB and transmit the power over the cold LV power cables, as shown in 
Figure~\ref{fig:pdune-wib}.

 The WIB receives and decodes the timing system clock and commands with an Analog 
Devices ADN2184 clock/data separator. It further processes the clock with a Silicon
Labs Si5344 jitter cleaner before it forwards the system clock and timing signals to 
the FEMBs, including the 100~MHz system clock. The Si5344 can also provide a local 
100~MHz clock, allowing the WIB to function as a complete standalone CE readout chain,
providing clock and LV power to all the FEMBs and transmitting the data to the 
control computer via the GbE UDP link. In case the 100~MHz clock from the WIB is 
lost or corrupted before arriving at the FEMBs, two commercial IDT FXO-LC736-100 
100~MHz oscillators are assembled on each FPGA mezzanine providing a backup clock 
source operating at cryogenic temperature. \\
The WIB receives the four 1.25~Gbps data links from each FEMB simultaneously, on 
four separate differential LVDS links. Each data link is processed onboard the WIB
by a Maxim Integrated MAX3000 equalizer. The WIB reorganizes and transmits the TPC data 
in the WIB FPGA, which can provide transceiver speeds up to 10.3125~Gbps, to 
the DAQ system over fiber-optical links. The CE diagnostic readout using the local WIB 
power, clock, and GbE link without requiring either the DAQ or timing systems was a 
crucial tool during the ProtoDUNE-SP CE installation and checkout tests described in 
Section~\ref{sec:install}.

\section{Production Electronics Quality Control}\label{sec:qaqc}

The large number of ProtoDUNE-SP TPC cryogenic electronics channels required a large-scale 
production electronics Quality Control (QC) effort. A set of dedicated test stands were implemented 
at BNL to evaluate CE components as part of this effort.
Specialized test hardware and associated software was developed for different components 
with the intention of automating most of the testing process. These test stands validated 
individual components, such as ASICs, as well as fully assembled FEMBs at room and cryogenic 
temperature to speed up production and improve reliability.

\subsection{Test Stand Software and Data Management}

A Python-based software framework was implemented to automate component testing and evaluation 
for production QC~\cite{femb_python}. Individual test stands used a Python script to coordinate 
the entire testing process for a specific type of component. These scripts controlled and 
configured test hardware, recorded relevant data and determined if the component under test met 
performance requirements.
The test software provided a simple GUI interface to allow operators to record key information 
and initiate the test process for a set of components. Recorded data was archived in a standard 
format, with key results stored in JSON format~\cite{json} for easy parsing.
Test software version control and the use of a git repository was enforced to ensure production 
testing was reproducible and documented~\cite{femb_doc}.
The Sumatra project tracking tool was integrated into the test software, allowing consistent 
logging and archiving of all production tests and their outcomes~\cite{sumatra_tool}. \\
Each test stand used a commercial personal computer (PC) with 5TB of disk space to run the test 
software and coordinate testing. All test hardware was controlled by either an FEMB FPGA mezzanine 
or a WIB, both communicating via the UDP Ethernet interface that was used
for real-time diagnostic readout, discussed in Section~\ref{wie}.
Additionally PC USB interfaces were used to control power supplies and function generators.
A "server'' PC controlled test stand PC configuration through a local network, archived data, and 
generated summary information and plots used to monitor overall production testing progress.
The server possessed 10TB of disk space, sufficient for the data recorded for production tests.

\subsection{Oscillator Tests}
\label{osc-test}

\begin{figure}[h!]
	\centering
	\includegraphics[scale=0.4]{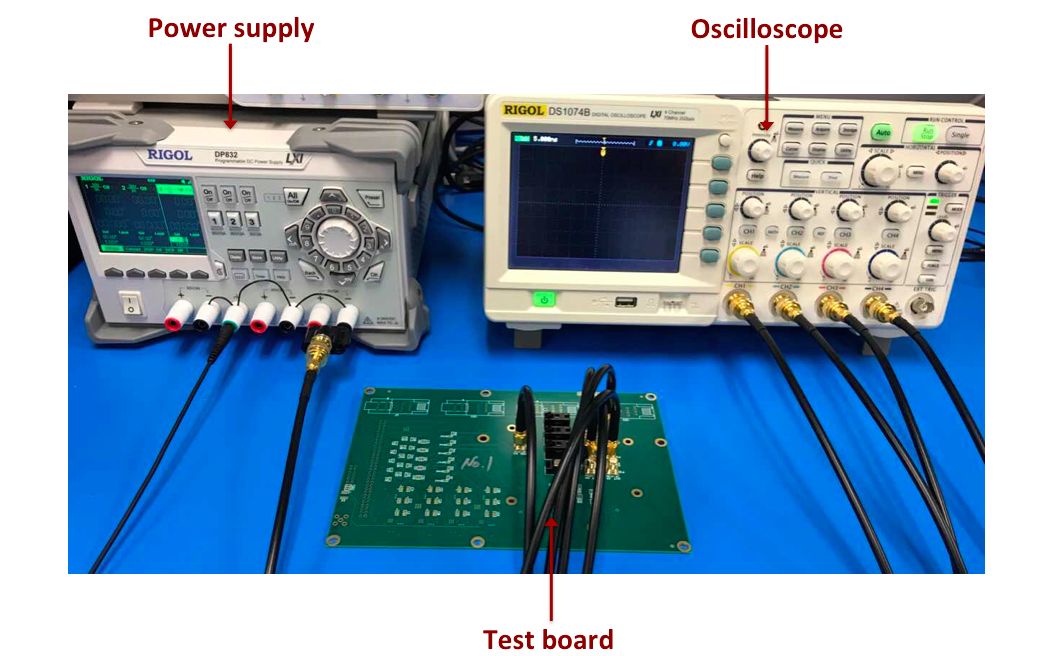}
	\caption{Quad-socket oscillator test board with test stand power supply and oscilloscope. 
		The oscillators are placed in the four onboard sockets, and power and signals are transmitted 
		through SMA cables.}
	\label{fig:qaqc_osc_testboard}
\end{figure}

Two 100~\si{MHz} oscillators are assembled on the FEMB FPGA mezzanine as discussed in 
Section~\ref{ce}. Each oscillator was tested in LN2 (77~K) to identify components likely 
to fail during normal cryogenic operation before assembly on the mezzanine. 
A dedicated test board allowed four oscillators 
to be tested simultaneously and is shown in Figure~\ref{fig:qaqc_osc_testboard}.
Oscillators were placed in individual test sockets by the test stand operator, with 
power and clock signals transmitted over SMA cables.\\
The test procedure required the operator to submerge the test board and each oscillator 
three times in LN2. The three separate cryogenic cycles were necessary to determine whether 
the oscillators could tolerate repeated cooldowns to cryogenic temperature.
While submerged the test script controlled a voltage supply to power the oscillators 
on-off 100 times. Each time the oscillators were turned on the clock frequency was measured 
by an oscilloscope and analyzed to check if consistent with 100~MHz.
Oscillators that failed to output the correct frequency on any power cycle during 
any of the three thernal cycles were rejected.\\
During production testing 700 oscillators were tested by the test stand in total, 
of which 450 were accepted for installation. The most common oscillator failure was 
failing to power on correctly when immersed in LN2, while a smaller number produced 
noisy output. Generally oscillators failed immediately when first powered on at cryogenic temperature.
The additional power cycles and immersions were intended to demonstrate that the oscillators that
passed the tests were likely to work through the expected operation period of the detector.

\subsection{Flash Memory Tests}

\begin{figure}[h!]
\centering
\includegraphics[scale=0.4]{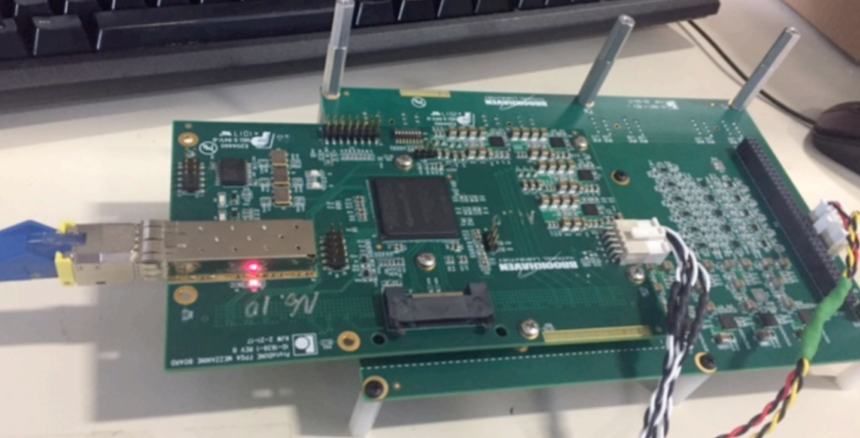}
\caption{Quad-socket flash memory test board, controlled by an FEMB FPGA mezzanine
with UDP Ethernet communication to the test PC.}
\label{fig:qaqc_flash_testboard}
\end{figure}

 One flash memory device is assembled on the FEMB FPGA mezzanine as discussed in 
Section~\ref{ce}. These devices were tested cryogenically to ensure they still functioned 
correctly before assembly on the mezzanine. A quad-socket test board allowed four 
devices to be tested simultaneously and is shown in Figure \ref{fig:qaqc_flash_testboard}.
This test board used an FEMB FPGA mezzanine to control the flash memory erase, 
read, and write operations, and communicate to the test PC via UDP Ethernet. \\
During a test the flash memory devices were inserted into the test board sockets by an operator, and then immersed in LN2 while powered off. Care was required to immerse 
the flash memory chips and not the Ethernet interface transceiver, which would not 
function correctly at low temperatures. After immersion in LN2 the test script was 
initiated and the flash memory erase, write, and read back functions were tested 
in sequence. Erase commands needed to complete within 180 seconds to be identified as 
successful, and a series of flash memory writes needed to be successfully read back 
within three attempts.\\
860 devices were tested in the test stand, of which 190 were accepted for installation.
Typical failures were for the flash memory to never report the erase command as successful, 
or for incorrect data read back following a write command. The high failure rate is not 
unexpected as the devices were not designed to operate cryogenically. 
Flash memory for firmware storage will be eliminated from the eventual DUNE design as it will replace the FPGA with a custom cold control ASIC, similar in concept to the FE and ADC ASICs operating in ProtoDUNE~\cite{dune-idr-v2}.
For the final detector, the FPGA will be replaced by a low power clock and control ASIC, eliminating the need for flash memory at cryogenic temperatures \cite{dune-tdr}.

\subsection{FE ASIC Tests}

960 FE ASICs are needed for ProtoDUNE-SP. By the end of the QC process 1192
validated FE ASICs were assembled onto production FEMBs, including spares.

\subsubsection{Warm Screening}
\label{fe-warm}

\begin{figure}[h!]
\centering
\includegraphics[scale=0.5]{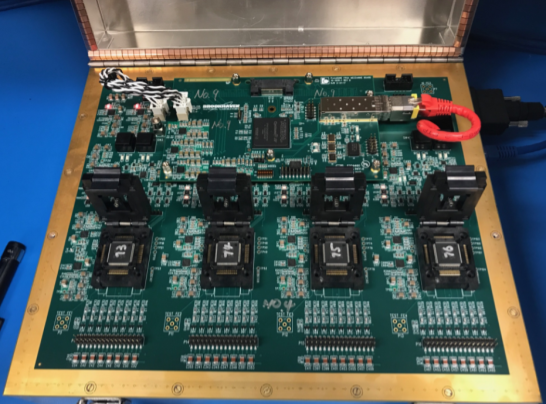}
\caption{Quad-socket FE ASIC test board for room temperature tests. The test
board is contained in a stainless steel box, which was closed by the test
operator before initiating the test scripts to provide a Faraday cage during
ASIC characterization.}
\label{fig:qaqc_preamp_rt_testboard}
\end{figure}

\begin{figure}[h!]
\centering
\includegraphics[scale=0.2]{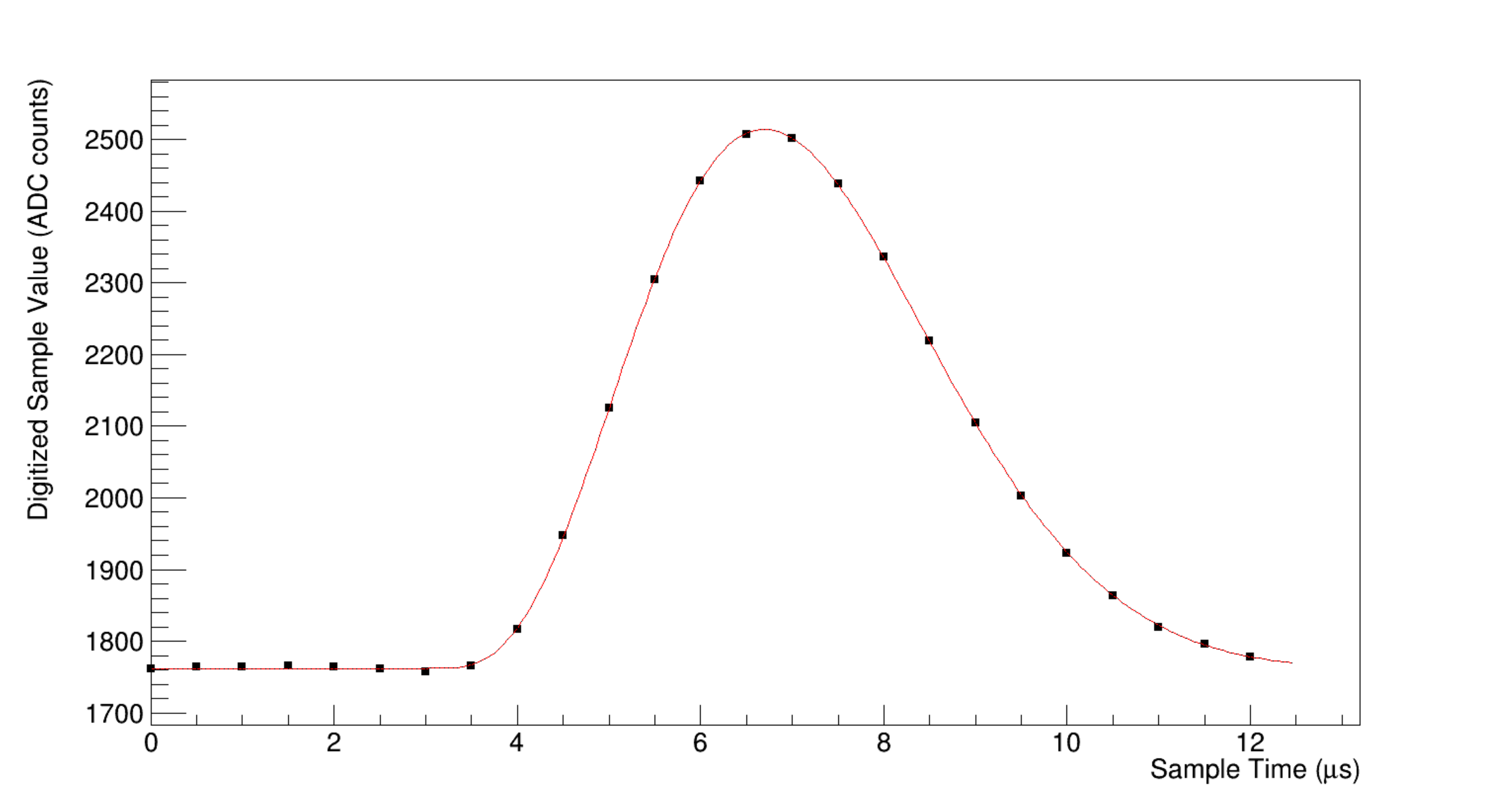}
\caption{An example calibration pulse in the digitzed channel waveform, 
with a fit to the ideal electronics response overlaid for comparison.}
\label{fig:qaqc_exampleCalibPulse}
\end{figure}

Each FE ASIC was tested at room temperature to avoid assembling FEMBs with 
defective ASICs. A quad-socket FE ASIC test board was developed to test up to four 
ASICs at a time, as shown in Figure~\ref{fig:qaqc_preamp_rt_testboard}.
An FEMB FPGA mezzanine controlled individual ASIC power supply and configuration.
Commercial Linear Technology LTC2314 14-bit ADCs digitized and recorded each ASIC 
channel's amplifier output, sending the digital serialized data to the FPGA which 
in turn transmitted it via UDP Ethernet to the test PC. An onboard DAC injects 
square wave signals into the FE ASIC injection capacitor connected to individual 
FE channel inputs (described in Section~\ref{ce}).
The amplifier electronic response was characterized using the resulting pulses 
in the digitized waveform, as shown in Figure~\ref{fig:qaqc_exampleCalibPulse}.

 To protect the FE ASICs from electrostatic discharge (ESD), operators wore ESD 
bracelets grounding them to a fixed reference in the lab. Status LEDs on the 
test boards indicated individual socket power status, reducing the chance of 
the operator damaging an ASIC by placing it in the test socket while the socket 
was powered. The test stand script GUI also reduced the chance of damage 
by directing the operator to place ASICs into the test board when the 
sockets were unpowered. \\
After the operator placed up to four FE ASICs in the test board and initiated 
the test script, each ASIC was power cycled, configured, and data recorded for
multiple FE ASIC configurations.
Individual channel electronic noise was measured with the RMS of the baseline.
Channel gains were measured by varying the magnitude of the DAC driving the 
injection capacitor and fitting the proportional change in average calibration 
pulse height measurements.
This data was analyzed and ASICs rejected if any channel noise and gain 
measurements were not within acceptable range. These measurements are done 
for all possible ASIC gain and shaping times, and recording data with pulses 
provided by the internal FE ASIC or external test board DAC.
Different leakage current settings, AC vs DC coupling, and buffer amplifiers 
were also tested.
The full set of tests are defined in Table~\ref{table:feasic_warm}.
1850 FE ASICs were tested at room temperature, with 103 rejected.
FE ASICs were generally rejected due to noise or gain measurements outside the acceptable ranges.
These ranges were defined as three standard deviations away from the mean of the parameter distribution as measured with a representatively large number of FE ASICs.
Some ASICs were also rejected due to individual channels failing to operate correctly.

\begin{table}[h!]
\centering
\begin{tabular}{|c|c|} 
 \hline
 Test Type & Number of Tests  \\
 \hline
 ASIC configuration scan with internal ASIC DAC & 32  \\ 
 ASIC configuration scan with FPGA DAC  & 4  \\ 
 Alternative leakage current setting scan & 3  \\ 
 Output buffer test & 1  \\ 
 AC output test & 1  \\ 
 \hline
\end{tabular}
\caption{Summary of pre-amplier ASIC test process and associated measurements.}
\label{table:feasic_warm}
\end{table}

\subsubsection{Cold Screening}
\label{fe-cold}

\begin{figure}[h!]
\centering
\includegraphics[scale=0.8]{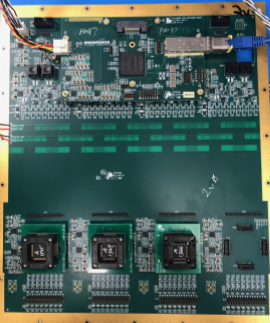}
\caption{Quad-socket FE ASIC test board for cryogenic testing with three of 
four ASIC sockets installed. The additional distance between the ASICs and 
digital electronics improved ADC reliablity by preventing them from becoming 
too cold to operate correctly during a cryogenic test.}
\label{fig:qaqc_preamp_ct_testboard}
\end{figure}

The production Analog Motherboards were assembled in batches of $\sim$25, to 
make the 20 plus spare FEMBs needed to complete a full APA module. For
the Analog Motherboards for the first 4 APAs, it was found that $\sim$5\%
of the FE had to be replaced after cryogenic testing, largely due to the 
collection-mode baseline 
drop-out issue discussed in Section~\ref{ce} or ESD damage during assembly. 
To prevent this rework, FE ASICs for the FEMBs for APAs 5 and 6 were also 
pre-screened at cryogenic temperature. \\
A dedicated ``stretched'' quad-socket cryogenic test board, 
shown in Figure~\ref{fig:qaqc_preamp_ct_testboard}, was developed 
with extended traces from the FE ASIC sockets to the LTC2314 ADCs and digital
control components. This reduced the chance of the commercial ADCs becoming 
too cold (around -50 $^{\circ}$C) during the test to operate correctly. These cryogenic tests 
determine if the FE ASIC collection channel baseline, nominally at 200~mV, has
not dropped below 100~mV due to the packaging stress, so that the channel is safely above the baseline change for good channel performance and could observe the injection capacitor pulses.
320 FE ASICs are needed for APA5-6 FEMBs. 550 ASICs were cryogenically tested, 
and 368 accepted. The majority of the ASICs initially rejected were due to failures
of the test sockets; upon retest of the rejected FE ASICs a failure rate of 
$\sim$4\% was observed, consistent with the FE failure rate when cryogenically
testing the FEMBs. \\
The commercial test sockets were not designed to operate at cryogenic temperature and 
would eventually become unable to make stable low-impedance contact to the ASIC pins due to socket deformation 
from repeated thermal cycles.

\subsection{ADC ASIC Tests and Selection} 
\label{adc-test}

\begin{figure}[h!]
\centering
\includegraphics[scale=0.4]{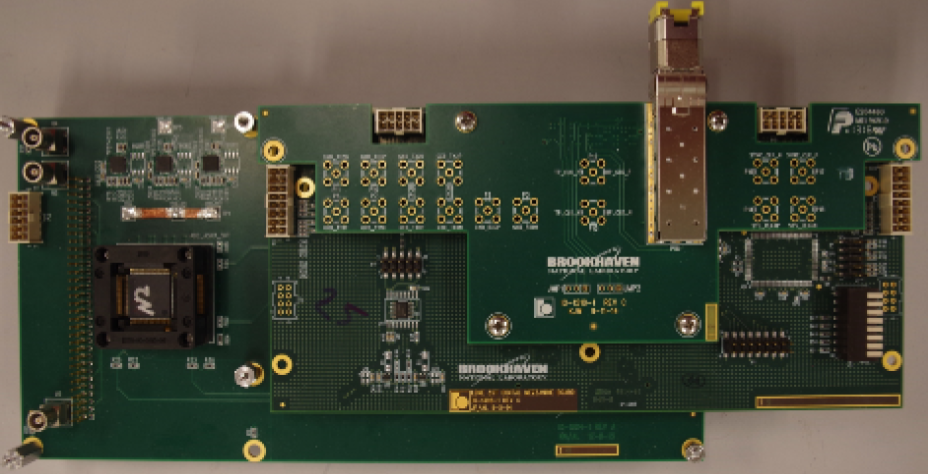}
\caption{Single-socket ADC ASIC test board, with ADC ASIC installed in socket.}
\label{fig:qaqc_adc_testboard}
\end{figure}

To optimize detector performance, a large number of ADC ASICs were tested at 
room and cryogenic temperature. The best performing ASICs were then selected 
for assembly onto the Analog Motherboards.
A single-socket ADC ASIC test board was designed to test one ASIC at a time, 
as shown in Figure~\ref{fig:qaqc_adc_testboard}. An FEMB FPGA mezzanine 
controls the ADC ASIC and reads out the two serial data streams, which it 
sends to the test PC via the UDP Ethernet interface. A function generator 
controlled by the Python test script provides input signals through a 
LEMO cable to the test board. 

\begin{figure}[h!]
\centering
\includegraphics[scale=0.4]{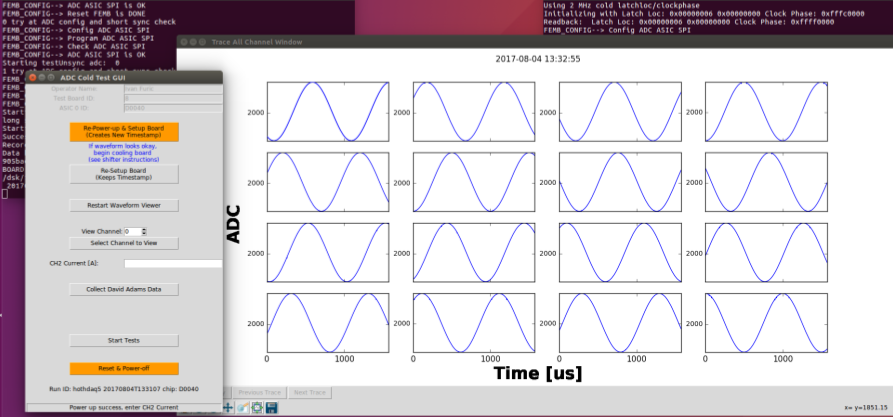}
\caption{Single-socket ADC test GUI and sinusoidal waveform display, indicating 
that the ADC readout is working correctly at room temperature prior to submersion 
in LN2.}
\label{fig:adcTestSineWaveDisplay}
\end{figure}

\begin{figure}[h!]
\centering
\includegraphics[scale=0.62]{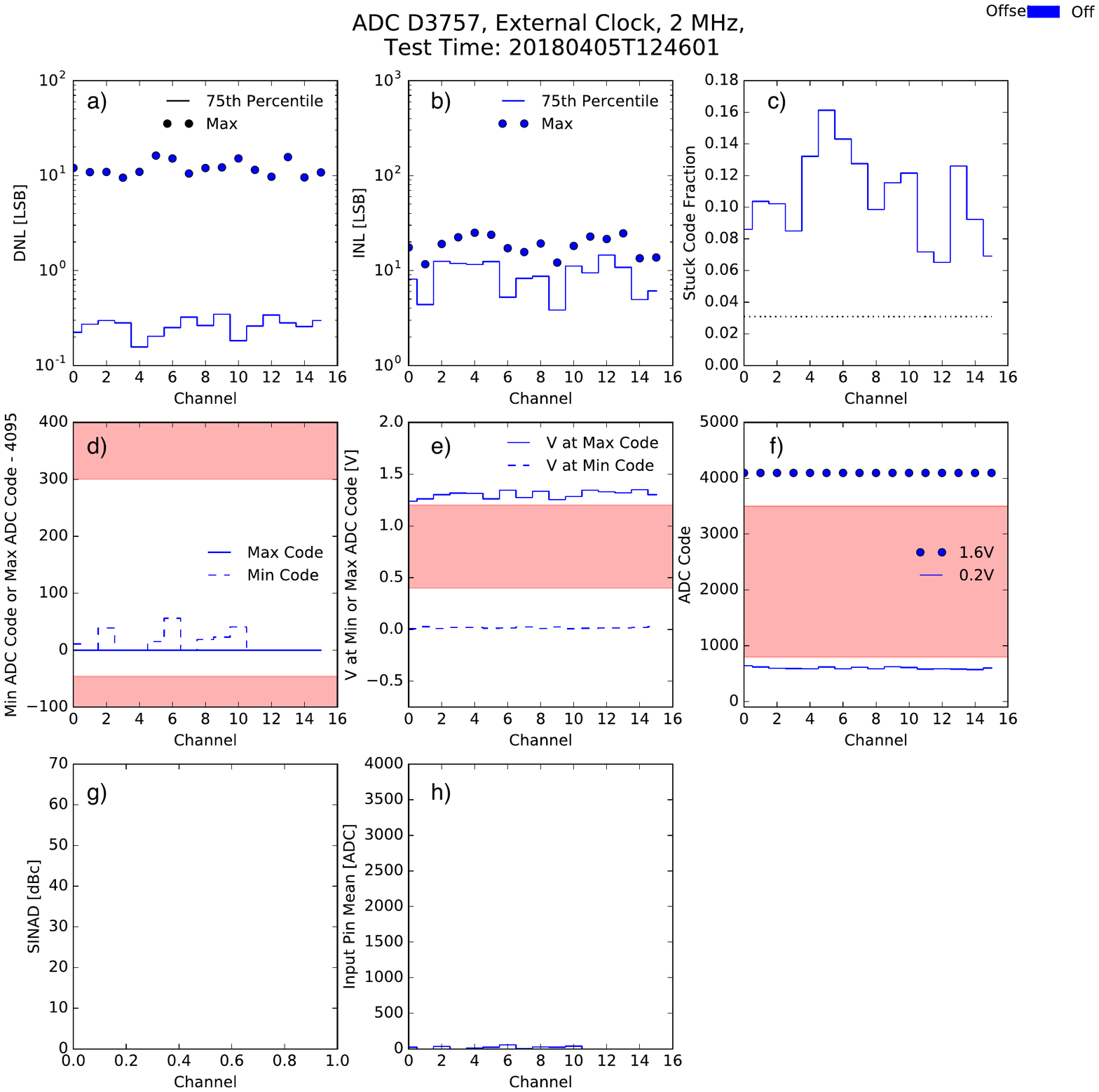}
\caption{Single-socket ADC test summary plot, showing the results of the cryogenic 
temperature ADC ASIC functionality and performance tests. Plotted results show 
channel-specific measurements of a) DNL, b) INL, c) fraction of samples with the 
stuck code error, d) minimum and maximum ADC codes within the ADC linear region, 
e) input voltage dynamic range measurement, f) average digitized sample code for 
0.2~V and 1.6~V input voltages, g) signal-to-noise and distortion ratio, h) average 
digitized sample code for disconnected ADC input. One summary plot was produced for 
each of the four ADC operating modes evaluated during the cryogenic test. \\
The pink regions represent areas in which the ADC performance is poor and that ASIC would be rejected by the test operator.} 
\label{fig:adcTestSummaryPlot}
\end{figure}

 To test a single ADC ASIC the operator places it into the chip socket while 
the test board is powered off. The operator follows the ESD protection
protocol described in Section~\ref{fe-warm}.
At this point the warm testing process is initiated through the GUI, which 
checks for basic functionality including whether the ASIC can be powered on, 
configured and the digital data stream synchronized correctly with respect 
to the FPGA digital logic. If this is successful a sinusoidal wave signal 
produced by the function generator is injected into all ADC ASIC channels and 
the resulting digitized waveforms displayed in real-time in a GUI window as 
shown in Figure~\ref{fig:adcTestSineWaveDisplay}.
This waveform display provides visual confirmation to the operator that the 
test setup is working and the ADC ASIC is functioning correctly at room 
temperature. At this point the ADC ASIC is submerged into LN2 
while the operator observes the real-time sinusoidal waveforms.
Care was taken to not immerse the Ethernet interface transceiver.
The real-time waveform display provided crucial feedback to the operator 
during the immersion process as to whether the ADC channels functioned 
properly: if a channel failed, the test was ended before the test script was 
launched. \\
Once fully submerged the operator started the cryogenic ADC ASIC test script,
the stages of which are:
\begin{itemize}
\item Injecting a long ramp signal into the ADC channel inputs with magnitude 
sufficient to cover the full ADC dynamic range. The digitized waveform of the 
long ramp signal was used to estimate ADC linearity and other parameters, 
such as input voltage range and including the sticky codes discussed in 
Section~\ref{ce}, from which an overall ADC quality metric was derived.
\item A series of functionality tests identified ADCs that failed to operate 
correctly cryogenically, including verifying the ASIC could be power cycled, 
configured, and record data at cryogenic temperature for all 16 input channels.
\item Measurements of ADC integral and differential nonlinearity (INL and DNL) 
and dynamic range were also used to reject underperforming ASICs.
\item Disconnected or malfunctioning channels were also detected, as were channels 
experiencing a higher than average incidence of the sticky code error.
\end{itemize} 
These tests were performed for four sets of ADC ASIC settings, where the source 
of ADC clock and control signals was switched between the ADC ASIC internal logic 
and logic provided by the FPGA, and also switching between 1\si{MHz} and 2\si{MHz} 
sampling rates. After these tests finished summary plots were automatically 
generated for the operator to review as shown in Figure~\ref{fig:adcTestSummaryPlot} 
indicating whether the ADC ASIC had been rejected \footnote{ The 12-bit ADC is expected to have bits counting 0-4095 available, however the protoDUNE ADC issues limited the available bits, so this figure shows deviations from min code (0) and max code (4095).\\ 
	These summary plots are shown without changes from the format they were displayed to the test operator during production testing.}.
Using this procedure 3680 ADC ASICs were tested, of which 271 where rejected 
outright. The remaining ADC ASICs were ranked in terms of performance based on the 
long ramp signal data, and the 1192 ($\sim$30\%) best-performing ASICs were 
selected for assembly. ADC ASICs were typically rejected for an inability to 
correctly synchronize their output serial data stream with the FPGA digital logic.
Poor performance or malfunction of one or more channels were also reasons for 
rejecting ADCs.

\subsection{Warm Interface Electronics Tests}

Dedicated test stands were not constructed for the warm interface electronics components 
(WIB, PTB, PTC, and flange PCB), due to the relatively low number of components and the less 
challenging room temperature operating requirement. However the components used in 
ProtoDUNE-SP also underwent a series of QC tests at BNL after assembly to ensure they 
functioned correctly when integrated and reading out FEMBs. Additional tests were 
performed during installation as described in Section~\ref{sec:install}.

\subsection{FEMB and CE Box Tests}
\label{ce-box}

\begin{figure}[h!]
\centering
\includegraphics[scale=0.4]{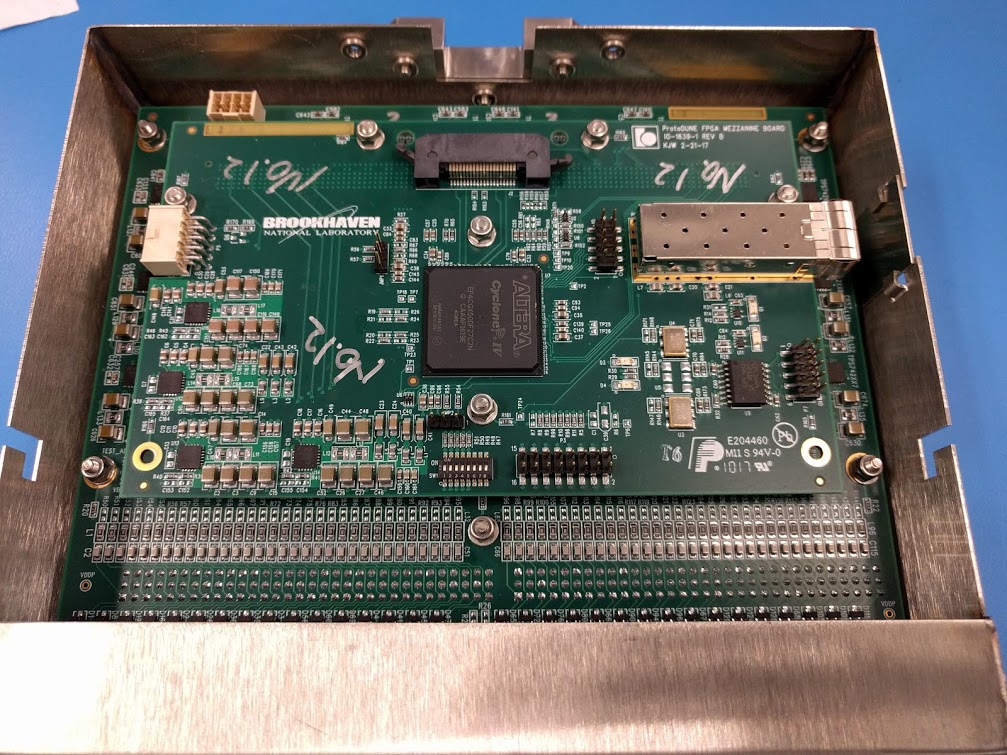}
\caption{Assembled FEMB integrated into a cold electronic box, prior to production testing.}
\label{fig:qaqc_fembInColdBox}
\end{figure}

Following individual component testing, FEMBs are assembled with components that passed 
previous tests. 120 FEMBs are needed to fully instrument all ProtoDUNE-SP APAs; including
spares, 149 FEMBs were assembled in the production run.
Assembled FEMBs then underwent pre-testing to verify basic functionality of a single FPGA along with multiple ASICs, and identify whether any rework was required. Successfully pre-tested FEMBs were then mounted within a "CE box'' as shown in Figure~\ref{fig:qaqc_fembInColdBox}, and 
fitted with a cold cable bundle and input pin adapter. The CE Box provides mechanical
support of the cables, protection of the FEMB components, and hardware for attaching
to the APA frame. This assembly was then tested as a unit at room and cryogenic temperature 
to fully evaluate functionality and performance. 
These tests were performed using the WIB UDP Ethernet readout used for real-time diagnostic commissioning of the detector, validating the complete CE readout chain from FE input to WIB output for every CE Box and set of cold cables.

\begin{figure}[h!]
\centering
\includegraphics[scale=0.4]{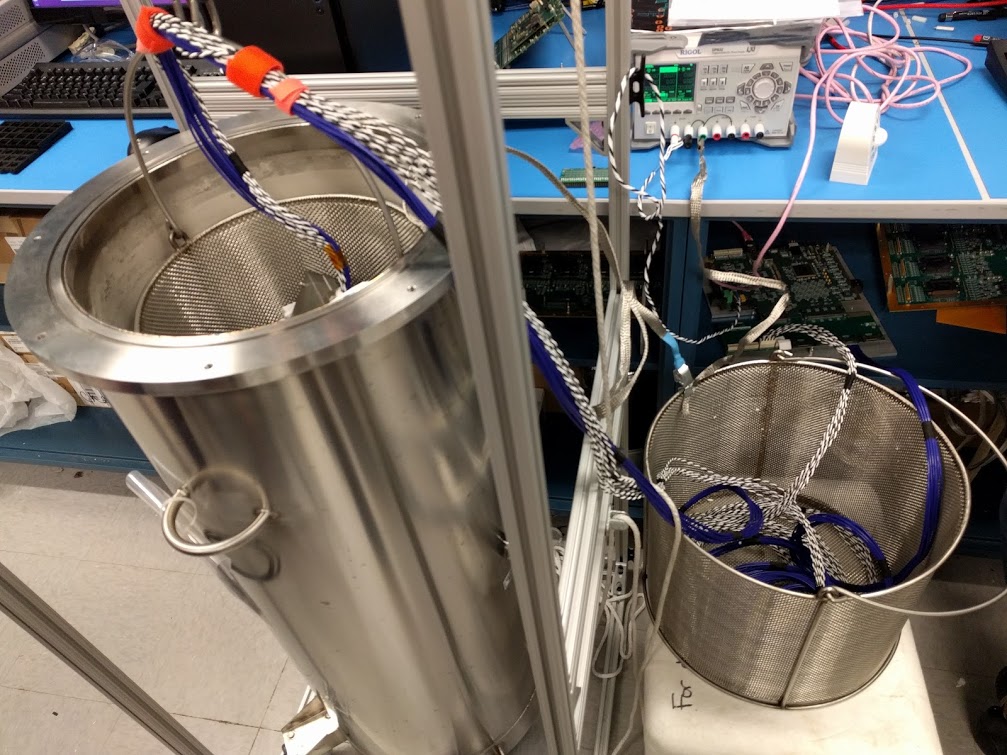}
\caption{The FEMB production test stand, showing LN2 dewar, mesh bucket containing the CE Box
inside the dewar, cold cables and WIB board. The code cables are held inside another wire mesh 
bucket outside the dewar during the test to minimize electronic noise pickup.}
\label{fig:qaqc_FEMBTestStand}
\end{figure}

 The initial FEMB test stand setup is shown in Figure~\ref{fig:qaqc_FEMBTestStand}.
The operator installed the CE Box in a wire mesh bucket that provided shielding and 
support during testing, and then initiated room temperature and cryogenic tests through a GUI.
The room temperature and cryogenic tests were identical aside from different configurations 
needed to syncronize the ADCs to the FPGA for room temperature and  cryogenic operation as 
discussed in Section~\ref{adc-test}. \\
Functionality tests included verifying the ability to power up and configure correctly, 
as well as checks that the voltage current draws on the power lines from the WIB to the FEMB are normal. \\ Semi-realistic ENC measurements were obtained by means of two "toy TPC'' PCBs that 
provide 150~\si{pF} input capacitance for every channel, which simulates the input
capacitance expected for a DUNE 7.5~meter APA wire. \\
Channel gain measurements were obtained similarly to the FE ASIC tests discussed in 
Section~\ref{fe-warm}, where both the internal FE and FPGA DACs were connected to the FE
ASIC injection capacitor and varied in size while the corresponding changes in average 
calibration pulse heights were measured. The resulting channel ENC and gain measurements 
obtained with different ASIC configurations were used to reject FEMBs with 
malfunctioning components or unacceptable performance.

\begin{figure}[h!]
\centering
\includegraphics[scale=0.6]{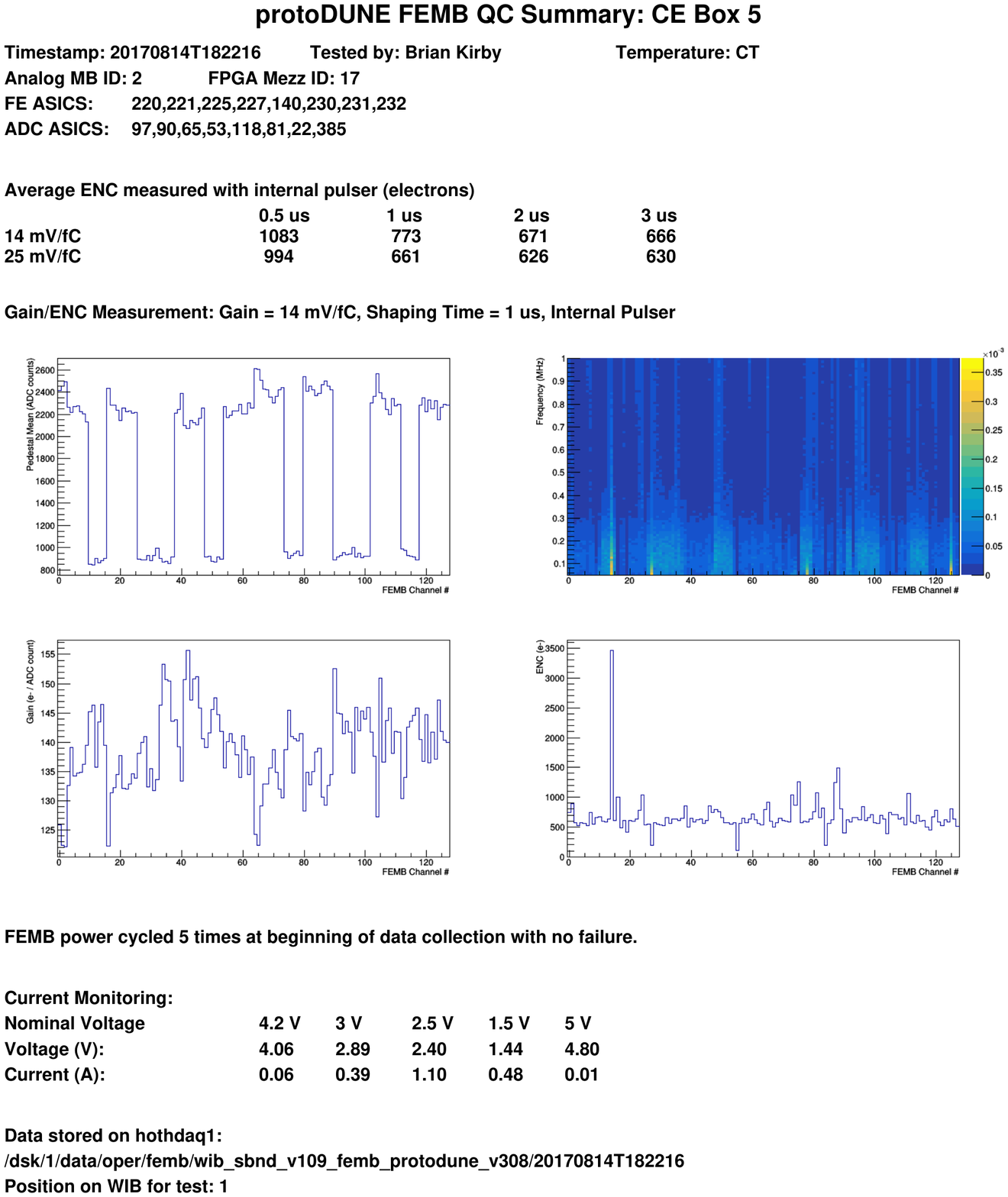}
\caption{Example FEMB QC test summary document, which summarizes results of the power cycle test, 
current draw, and average ENC measurements used to evaluate FEMB performance.}
\label{fig:fembSummaryDocument}
\end{figure}

After the room temperature tests, the CE Box was slowly immersed into LN2 and the test script 
initiated again for the cryogenic tests. At the end of each of the room temperature and cryogenic 
test process a summary document was automatically generated as shown in 
Figure~\ref{fig:fembSummaryDocument}. This document summarized the tests performed and board 
performance, and was used to identify and reject boards with malfunctioning channels, 
excessive electronic noise or failure to function correctly.

Of the 149 FEMB and CE Box production assemblies tested, $\sim$4\% required expert technician 
rework of the FEMB, and four were rejected due to unrepairable excessive 
ENC or poor performance in one or more channels. FEMBs that passed these final QC tests were 
delivered to CERN for acceptance testing and installation, 
as described in Section \ref{sec:install}.

\subsection{Cryogenic Test System}

\begin{figure}[h!]
\centering
\includegraphics[scale=0.5]{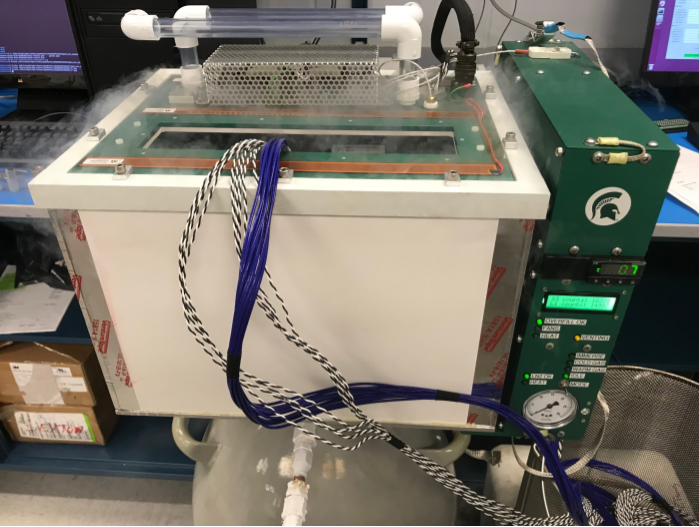}
\caption{A production CE Box held within the CTS for a quality control test. The CTS 
automates the LN2 immersion and warming back to room temperature processes.}
\label{fig:qaqc_fembInMSUDewar}
\end{figure}

 An automated Cryogenic Test System (CTS) was introduced to the FEMB quality control tests 
and used to test the final $\sim$50\% of the production CE Boxes. The CTS with a CE Box under 
test is shown in Figure~\ref{fig:qaqc_fembInMSUDewar}. In addition to testing CE Boxes, 
the CTS was designed to immerse the FE ASICs in the stretched quad-socket test board shown
in Figure~\ref{fig:qaqc_preamp_ct_testboard}. Two CTS were used to cold screen the FE ASICs
for APA5-6, as discussed in Section~\ref{fe-cold}. A total of three CTS were used for 
the final production QC tests at BNL. \\
The CTS automates the immersion of components in LN2 for cryogenic testing 
purposes, as well as the subsequent warming back to room temperature. A key 
improvement over the open-top dewar test stand is that the CTS flushes the 
chamber holding the component under test with cold nitrogen gas before 
immersion, removing any humidity that might condense on the component. Similarly 
after the QC test is complete the CTS warms the chamber in nitrogen gas, reducing 
the chance of water condensation and protecting the components under test. The CTS
did not require altering the production tests: when
the FE ASICs or FEMB were immersed in the CTS, exactly the same GUI-controlled
test software discussed in Sections~\ref{fe-cold} and~\ref{ce-box} was used to
run the production QC tests.
ENC measurements were $\sim$10\% higher for tests performed in the CTS over the 
open-top dewar, due to the operation of liquid level sensors in the CTS. This did
not affect identification of malfunctioning or under-performing FE ASICs or FEMBs.

\subsection{Summary and Lessons for Future LArTPC Experiments}

A summary of the QC test results for all components is shown in 
Table~\ref{table:qaqc_summaryTable}. The relatively low rejection rate of FEMBs was 
a major achievement and in large part due to pre-screening of individual components.
Reworking already assembled FEMBs to replace a failing component requires an expert 
technician and can damage other components; minimizing the rejection rate reduces 
this risk. \\
The ProtoDUNE-SP QC testing effort provided a great deal of experience that can be 
applied to future LArTPC electronic productions. One shortcoming in test design was 
the failure to test FE ASICs with an extended test signal to simulate a long ionization 
charge distribution. This is a common type of charge distribution within a LArTPC and 
the associated electronics response should have been evaluated.
Simplification of the GUI software and automation of the testing process and component 
pre-testing helped the QC test stands process the electronics production in a timely 
manner, allowing non-expert operators to be quickly trained and run the test stands in 
shifts, and is an approach that should be carried forward to future efforts.

\begin{table}[h!]
\centering
\begin{tabular}{|c|c|c|} 
 \hline
 Component & Number Tested & Rejection Rate  \\
 \hline
 100\si{MHz} Oscillators & 700 & 35.7\%  \\ 
 Flash Memory & 860 & 77.9\%  \\ 
 Pre-amplifier ASIC (warm) & 1850 & 5.6\%  \\ 
 Pre-amplifier ASIC (cold) & 168 & 4.2\%  \\ 
 ADC ASIC & 3680 & 14.6\%  \\ 
 FEMB & 149 & 2.7\%  \\ 
 \hline
\end{tabular}
\caption{Summary of ProtoDUNE-SP electronic component QC tests.}
\label{table:qaqc_summaryTable}
\end{table}


%
%
%


\section{Installation} \label{sec:install}

\begin{figure}[h!]
	\centering
	\includegraphics[width=1\linewidth]{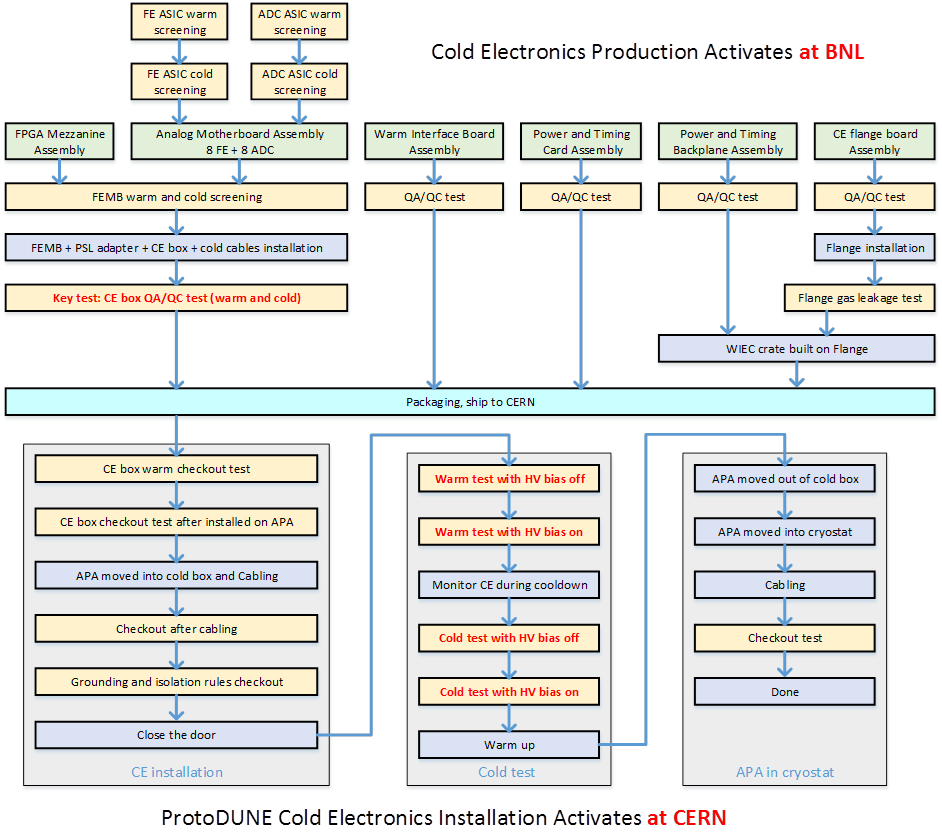}
	\caption{Flow chart showing the different stages for the cold electronics production at BNL and installation and activation at CERN.} 
	\label{fig:flow}
\end{figure}

The following flow chart (Figure \ref{fig:flow}) shows all the testing steps performed separately at BNL and CERN on all the different CE components.
In the following section will be described the installation and activation procedures carried out at CERN.
As shown in the plot, the installation at CERN included three different steps. 
The first one (CE installation) includes a set of checkout test performed on the CE boxes from their delivery at CERN to their installation on the APA. 
Those tests, along with the Cold Box test described in Section \ref{cold-box}, allowed to verify the functionality of all readout channels and the ENC levels before inserting the APAs into the cryostat. 
The last set of tests allowed to verify and maintain the electronics performance during the detector commissioning where the installation of other subsystem may affect the electronics functionality.

\subsection{Reception test and installation }
Once delivered to CERN, the CE Boxes were tested several times throughout the 
installation steps. A first reception test was performed at room temperature on 
all the CE Boxes before installation on the APA. To protect against ESD damage
to the FEMBs, the CE Boxes were shipped with the cold cables detached from the
electronics. Once the cold LV and data cables were re-assembled, the shorting 
caps were removed and the CE Box was connected to a test setup consisting of a readout PC connected via Ethernet to a WIB, 
and an adapter board simulating the WIEC connection. Two 150~pF toy TPCs, 
described in Section~\ref{ce-box}, were connected to the CE Box inputs to 
simulate the APA capacitance. \\
ENC was measured on all channels for all FE ASIC gain and shaping time settings. 
The pulse response of each FEMB was checked by injecting bipolar pulses from 
the FE ASIC calibration circuit described in Section~\ref{ce}. Through averaging 
calibration response over many pulses, functionality of all readout channels was 
verified.
CE Boxes were only accepted for APA installation if all 128 channels were functional
and observed ENC levels typical for room temperature operation: $1000-1500$~e$^-$.\\
The same test was repeated on each CE Box after the installation on the APA and 
on sets of four CE Boxes connected to the same WIB after cabling. CE Boxes 
with any loss of channel functionality after installation were removed from the
APA and replaced with a spare CE Box.
Several CE Box failures were observed at these stages: 3 at the reception test, 
one after the installation, and none after the cabling. In Table~\ref{Failure}, 
a detailed summary of CE Box failure at these and the following testing stages 
is shown. One further CE Box failure occurred during the detector filling which 
is described in Section~\ref{FEMB302}.\\
A data cable connector failure, which resulted
in loss of communication between the WIB and the CE Box, was the most common 
issue, especially during the Cold Box test at RT and during the cooldown described
in Section~\ref{cold-box}. Microscope analysis of a damaged FEMB performed at BNL 
revealed some micro-cracks on the connector welding to the FEMB FPGA mezzanine, 
causing the connector to partially separate from the PCB. Further testing at 
BNL confimed that this failure could be induced through a combination of mechanical 
and thermal stress. For this reason, a new data connector design for the next FEMB 
version is currently under review.

\begin{table}
	\centering
	
	\begin{tabular}[c]{c | c | c | c }
		\centering
		
		\textbf{APA number} & \textbf{CE Box ID} & \textbf{Failure description} 
                            & \textbf{Testing stage} \\
		\hline 
		
		& & &  \\
		1 & 9 &  1 dead FE channel at RT & QC test at BNL \\
		& 20 &  LV return wire cut during cabling on APA
		& Installation \\
		& 24 & 3 dead FE channels at RT & Installation \\
		\hline
		
		& & &  \\
		2 & 39 & Data cable connector failed during cooldown & Cold Box Test \\    
		\hline
		
		& & &  \\
		3 & 69 & 1 dead FE channel at RT & Reception test \\
		& 49 & Data cable connector failed at RT & Cold Box Test \\ 
		& 18 & Data cable connector failed at RT & Cold Box Test \\   
		& 22 & 1 FE ASIC (16 channels) failed during cooldown &
		Cold Box Test \\
		& 75 & Data cable connector failed at RT & Cold Box Test \\   
		\hline
		
		& & &  \\
		4 & 91 & 1 dead FE channel at RT & Reception test \\
		& 85 & Data cable connector failed during cold test & Cold Box Test \\   
		\hline
		
		& & &  \\
		5 & 122 & 2 links failed during warm test & Cold Box Test \\
		& 123 & Data cable connector failed at RT & Cold Box Test \\   
		& 106 & Data cable connector failed at RT & Cold Box Test \\  
		\hline
		
		& & &  \\
		6 & 112 & Data cable connector failed at RT & Cryostat checkout \\ 
				
	\end{tabular}
	\caption{Detailed summary of the CE Box failure at different testing stages 
                          during the TPC installation and commissioning. }
        \label{Failure}
\end{table}

\subsection{Cold Box Integration Test}
\label{cold-box} 

\begin{figure}[h!]
	\centering
	\includegraphics[width=0.6\linewidth]{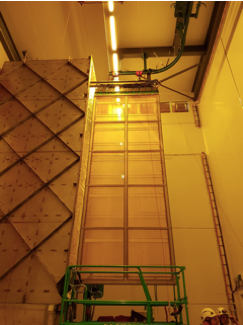}
	\caption{Picture of the Cold Box facility for APA integration test. In the picture, the APA-assembly (APA+CE system+PD system) is ready to be inserted into the cold box. On top of the APA you can also see 10 out of 20 installed CE boxes. }
	\label{fig:CB}
\end{figure}

\begin{figure}[h!]
\centering
\includegraphics[width=1\linewidth]{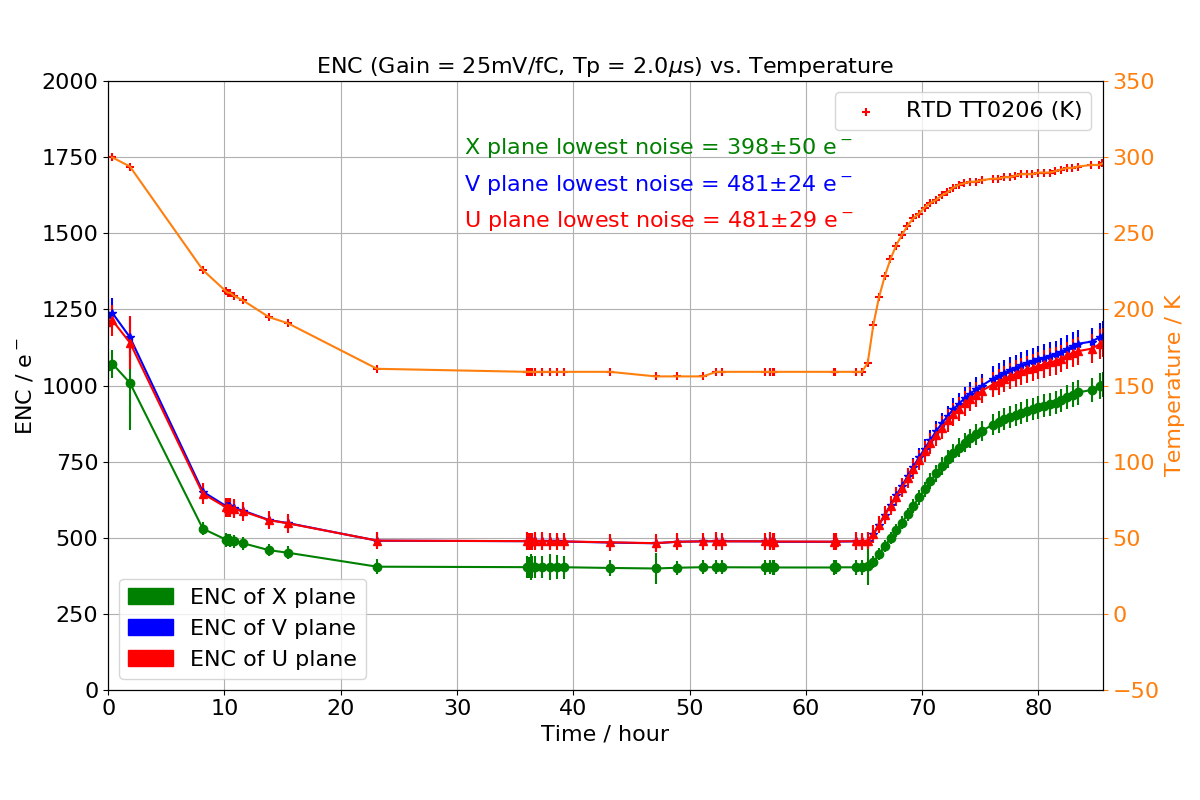}
\caption{ENC performance in electrons of the APA+CE during the Cold Box test as 
a function of time; red/blue are the collection (U/V) wires, green are the induction
(X) wires. The temperature is measured by RTD sensors installed on one side of the APA;
the orange curve is the RTD at the level nearest the CE Boxes.}
\label{fig:CB_noise}
\end{figure}

\begin{figure}[h!]
\centering
\includegraphics[width=1\linewidth]{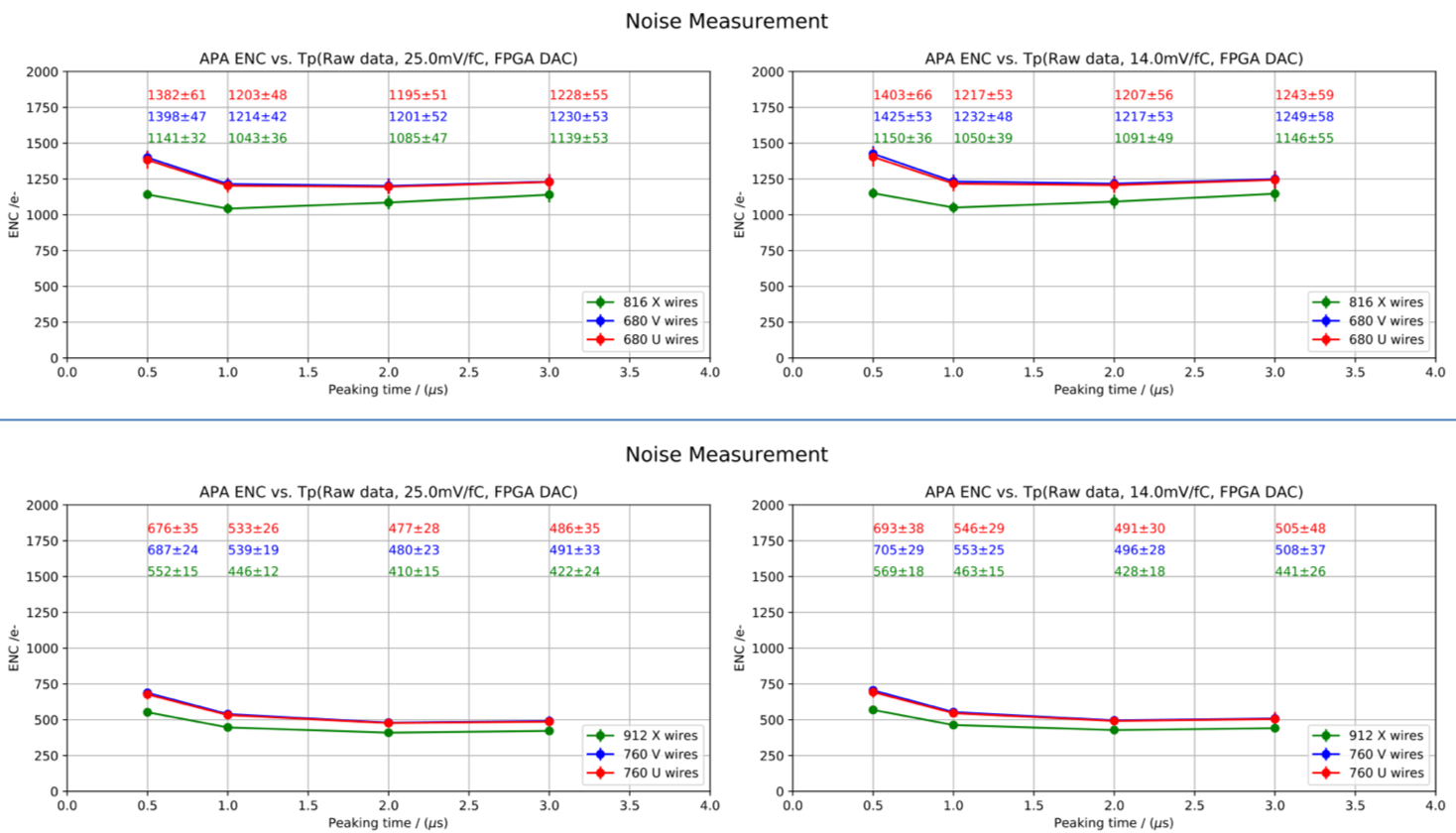}
\caption{Comparison between average ENC (e$~-$) at warm (\textit{top}) and cold (\textit{bottom}) 
temperature for two different gain values (25~mV/fC \textit{left}, 14~mV/fC \textit{right}): 
red/blue are the collection (U/V) wires, green are the induction (X) wires. The overall 
lower ENC for the X wires is due to the shorter wire length of 6~m compared to the $\sim$7.5~m
U/V wires.}
\label{fig:ENC_CB}
\end{figure}

In order to evaluate fully-instrumented APA performance at cryogenic temperature, 
a Cold Box was built for integration testing at CERN (Figure \ref{fig:CB}). The integration test included 
a full production signal feed-through assembly and WIEC, described in 
Section~\ref{wie}, containing production PTB, PTC, and five WIBs. This allows a 
vertical slice test of all APA wires, CE readout, and photon detectors (described 
in Chapter 2.7 of~\cite{protodune-sp-tdr}) on production APAs before insertion 
into the cryostat.
The CE readout through optical fibers from the WIBs allows a real-time study of 
the detector performance in Cold Box tests, independent from the full DAQ readout.
Seven temperature sensors installed on one side of the APA allows continuous 
monitoring of the temperature inside the Cold Box and the temperature gradient 
between the bottom and the top of the APA where the CE is installed.
The APAs were cooled down over an approximately 24~hour period to $\sim$150~K at the
CE by injecting cold nitrogen gas from the top of the cold box.\\
The ENC performance was evaluated first at room temperature and during all the cold 
test stages (cool down, stable cold temperature, and warm up). In 
Figure~\ref{fig:CB_noise} the ENC with FE gain of 25~mV/fC and shaping time 
2~$\mathrm{\mu}$s of the induction planes (blue and red) and 
collection plane (green) for APA2 are shown as a function of time; the temperature 
values corresponding to the several temperature sensors (orange) installed on one 
APA side are also shown. At room temperature, the ENC is around 1200~e$^{-}$ for the 
induction planes and 1100~e$^{-}$ for the collection plane.
At the lowest temperature achievable by the Cold Box facility, the ENC reaches a 
minimum value at $\sim$500~e$^{-}$ for the induction planes and $\sim$400~e$^{-}$ 
for the collection plane, matching to the ENC projection presented 
in~\cite{protodune-ce2} after correction for the wires in gaseous rather than
liquid nitrogen.
In Figure~\ref{fig:ENC_CB}, a more detailed comparison between the 
ENC at warm and cold temperature is shown for FE gain values 14 and 25~mV/fC 
as a function of shaping time, demonstrating the advantages of the CMOS 
technology at cryogenic temperatures.

\section{Commissioning} \label{sec:commission}

\begin{figure}[h!]
\centering
	\includegraphics[width=0.9\linewidth]{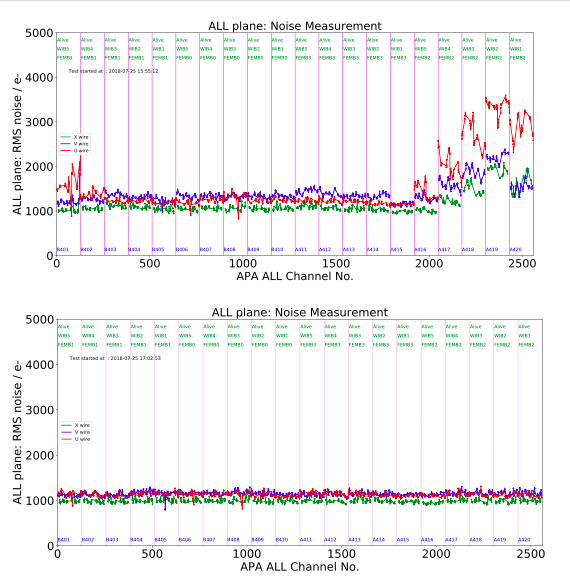}
\caption{Comparison between RMS on APA4 before (\textit{top}) and after (\textit{bottom}) 
the replacement of 12V power supplies for LEDs and cameras with a linear DC LV supply.}
\label{fig:Noise_APA4}
\end{figure}

In April 2018, the TPC installation was complete and all its components were 
positioned inside the cryostat and commissioning activities started in order 
to be ready for the first detector data beam run on September 2018. 
Commissioning of the detector included fixing the APA, CPA, and field cage 
end walls in their final position, installing the high voltage cup and 
feedthrough and all the cryo-instrumentation, cameras, LEDs, purity 
monitors, and temperature sensors. \\
Electronics commissioning comprised attaching the cold LV and data cables to
the CE signal feed-throughs, Warm Interface Electronics installation, installing 
the optical fibers and checking the CE-to-DAQ-system connection. Before starting 
the cabling at the WIEC, the functionality of all FEMB channels was checked again 
by using the stand-alone WIB test setup and identical metrics to those described
in Section~\ref{sec:install}. No variations were found from the last test before
moving the APAs inside the cryostat. 
Finally, as each set of four CE Boxes were cabled to the signal feed-throughs, 
the correspoding WIB was inserted in the WIEC and an identical set of tests
were run; after all cables and WIBs were installed, one FE channel was not 
responding to the FE calibration pulser at room temperature (0.0065\%).\\
Throughout the detector commissioning, the baseline, RMS, and calibration pulser 
response of the CE have been periodically monitored by the WIB diagnostic readout. 
A significant increase was observed in RMS after the installation of LEDs and 
cameras, especially on the closest APAs (APA4 and APA6). A series of tests on 
APA4 showed that the noise was induced by the 12V power supplies installed for 
the LEDs and cameras. Even without powering on the LEDs and cameras (supply on, 
setting 0V), the noise was still there transmitted to the detector ground by the 
return wire. The solution was to replace the 12V power supplies with a linear
DC LV supply. Figure~\ref{fig:Noise_APA4} shows the RMS observed on 
APA4 before and after the replacement of the 
power supply. In both cases, cameras were on during the test. However, after power supply replacement, 
two peaks at $\sim$500 and 600~kHz were still observable with a 
Fast Fourier Transform (FFT) analysis in the frequency domain due to LEDs and cameras. 
The remaining noise after replacement has been filtered 
at the offline data analysis stage. \\

 In August 2018, detector commissioning was completed and the cryogenic group started the purging and filling procedures. In September 2018, the detector was fully filled 
with LAr and ready for the activation procedure. 

\subsection{Electronics Performance}

Table~\ref{TPCchannels} shows a detailed summary of the TPC channels performance
measured in several tests during the detector activation. The last test is September 23 (2018), 
when the nominal cathode drift voltage value of -180~kV was reached and detector was 
fully operating for the first time. 

\begin{table}
	\centering
	\footnotesize 
	\begin{tabular}[c]{r | c | c | c | c | c | c | c | c | l }
		\centering

\textbf{Test ID} & \# 1 & \#5 & \#9  & \#11 & \#13 & \#15 & \#18 & \#35 & \\
\textbf{Date} & 9/13 & 9/14 & 9/16 & 9/16 & 9/17 & 9/19 & 9/20 & 9/23 &  \\
\textbf{HV status} & off & -120kV & off & off& off& off & -160kV & -180kV & \\

\textbf{Not Working Ch.} & 112 & 112& 112& 112& 112 & 0 & 0 & 0 & ADC sync error \\ 
 & 13 & 40 & 0 & 12 & 12 & 6 & 0 & 0 & FE with start-up\\
 & 0 & 3 & 2 & 2 & 2 & 2 & 2 & 4 & Inactive FE channels \\
 & 6 & 4 & 4 & 4 & 4 & 4 & 2 & 2 & FE gain < 180 e$^{-}$/ADC \\
 & 48 & 52 & 48 &  46 & 48 & 46 & 59 & 45 & Non-removable stuck code \\
 & 41 & 38 & 37 & 38 & 36 & 37 & 39 & 38 & Open connection \\
 & 2 & 0 & 0 & 0 & 0 & 0 & 1 & 3 &  ENC > 2000 e$^{-}$ \\
 & 295 & 348 & 334 & 330 & 292 & 318 & 405 & 386 & 2000 e$^{-}$ < ENC > 1000 e$^{-}$ \\
 & 446 & 466 & 451& 463 & 457 &442 & 655 & 627 &  1000 e$^{-}$ < ENC > 800 e$^{-}$ \\
\textbf{Good Ch. (ENC < 800 e$^{-}$)} & 14397 & 14297 & 14372 & 14353 & 14397 & 14377 & 14179 & 14259 & \\
\textbf{Active FE Ch.} & 15229 & 15201 & 15242 & 15230 & 15230 & 15220 & 15338 & 15354 & \\
\textbf{Active TPC Ch.} & 15188 & 15163 & 15205 & 15192 & 15194 & 15183 & 15299 & 15318 & \\
 
\end{tabular}
\caption{Performance of TPC channels measured throughout detector activation. The last column 
shows how CE failures have been classified. Failures are listed from high to low priority. } 
\label{TPCchannels}
\end{table}

\begin{figure}[h!]
\centering
\includegraphics[scale=0.6]{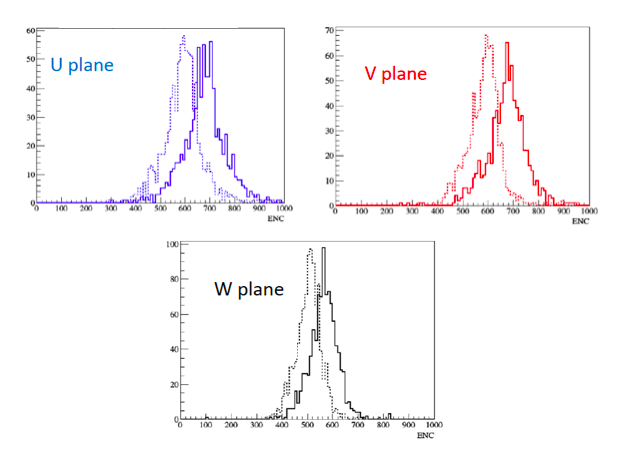}
\caption{Overall ENC observed on RUN 5102 on the three wires planes. For each plot, the ENC before (dashed line) 
and after (solid line) the offline noise filtering is shown.}
\label{fig:ENC_overall}
\end{figure}

\begin{figure}[h!]
	\centering
	\includegraphics[scale=0.4]{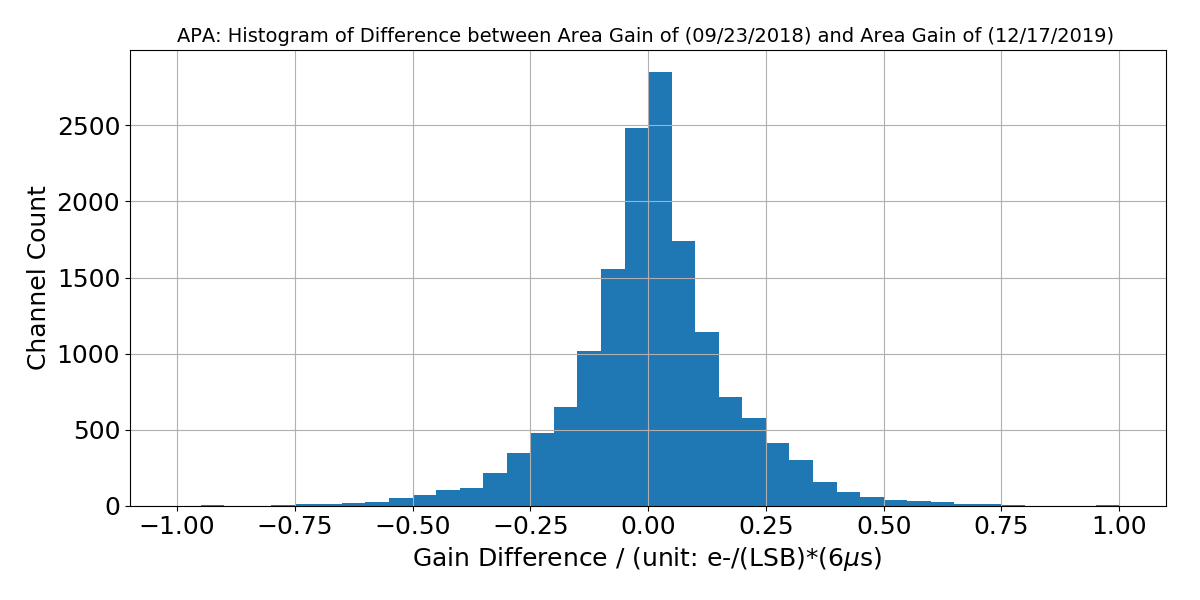}
	\caption{FEMB inverted gain difference between the beginning of the full-detector operations (Sep 2018) and one year later (Dec 2019)}
	\label{fig:APA_gain}
\end{figure}

\begin{figure}[h!]
	\centering
	\includegraphics[scale=0.4]{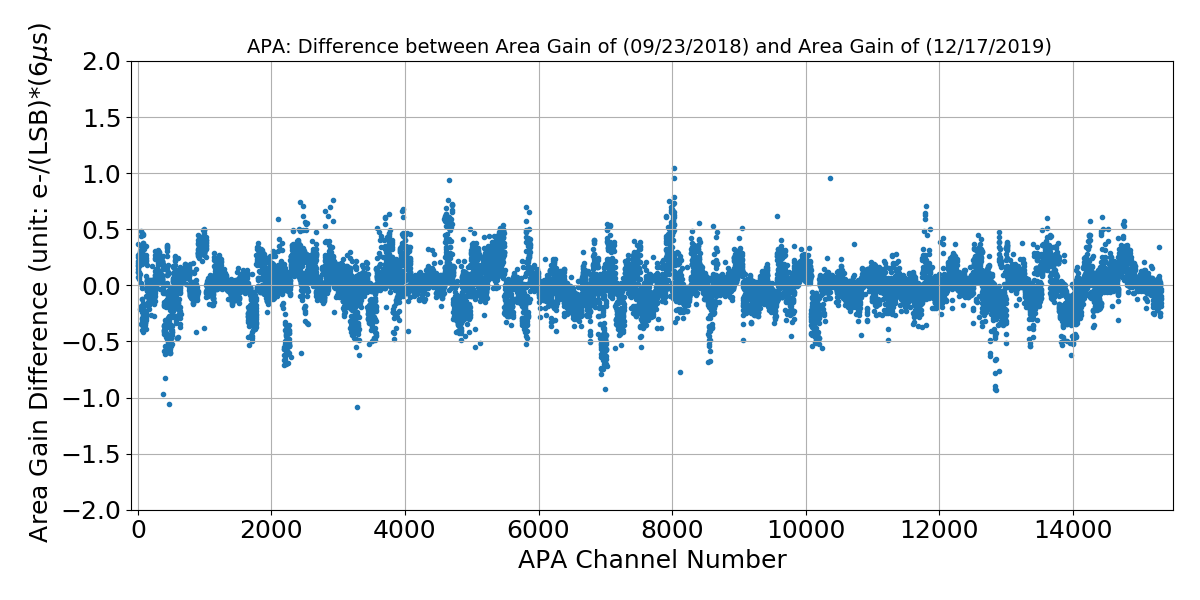}
	\caption{Channel by channel difference of the inverted gain between the beginning of the full-detector operations (Sep 2018) and one year later (Dec 2019)}
	\label{fig:APA_Hist_gain}
\end{figure}

 As of August 2019, after $\sim$7 weeks of beam data-taking and $\sim$8 months of cosmic rays 
data-taking, 42 channels are found to be non-operational, with 99.7\% of 15,360 TPC channels 
in total working properly. Of those 42 channels, 38 are missing or disconnected wire 
candidates and 4 are non-responsive to the FE calibration pulser.
93\% of the TPC channels are working with excellent noise performance (ENC<800e$^{-}$),
well below the CE design noise requirements for the DUNE FD. For the remaining 7\%, the abnormally 
high RMS has been mostly correlated in further studies at CERN to either the drift HV power supply 
operation or the TPC instrumentation, as already observed during the commissioning stage.
In Figure~\ref{fig:ENC_overall}, the overall ENC observed on a standard checkout run (RUN 5102) 
is shown for the three wires planes. \\
In Figure \ref{fig:APA_gain} is shown the FEMB inverted gain comparison 
between the beginning of the full-detector operations (Sep 2018) and one year later (Dec 2019). 
The measured gain shift is around $0.13\%$ and $0.4\%$RMS, with an excellent  agreement between 
2018 and 2019  (around $0.01\%$ shift with $0.3\%$ RMS). In Figure \ref{fig:APA_Hist_gain}, is shown 
the comparison channel by channel of the inverted gain between the same two runs.  

\subsection{Clock failure on FEMB302}
\label{FEMB302}

\begin{figure}[h!]
\centering
\includegraphics[scale=0.5]{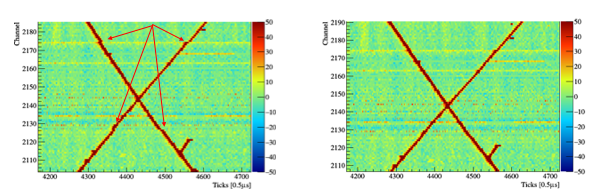}
\caption{Example of a particle track before (\textit{left}) and after (\textit{right}) the 
offline data filtering. Synchronization errors on the track (pointed out by the arrows 
on the left) have been fixed restoring the original track shape.}
\label{fig:Filtering}
\end{figure}

During detector filling, communication was lost to one CE Box on APA3. The failure mode
for this CE Box was similar enough to the ones in the Cold Box to suggest an FEMB data connector
failure. However, a detailed check with the WIB Ethernet diagnostic readout showed that the FEMB
link connection was still active, suggesting that only the system clock pin on the data
connector was actually broken causing a communication interruption between WIB and FEMB.
To recover the FEMB, a new firmware version for the FEMB FPGA that bypassed the system
clock and used the backup oscillators discussed in Sections~\ref{ce} and~\ref{osc-test} as
the FEMB clock was programmed over the backup JTAG links. The unavailability of the system clock 
caused the ADC digitization on all 128 FEMB channels to be asynchronous with the rest of the CE
system. To correct this, data filtering is performed offline using track candidates to correct
synchronization errors and restore the original shape of the track.
Figure~\ref{fig:Filtering} shows an example of a particle track before and after the offline correction.

\section{Conclusion} \label{sec:summary}

The ProtoDUNE-SP detector at the CERN Neutrino Platform has been a successful
validation of the Cold Electronics LArTPC readout system for the DUNE Far
Detector. The CE system contains all the electronics necessary to amplify, 
shape, digitize, and transmit the TPC wire data out of the cryostat to the 
DAQ system while operating at cryogenic temperatures ($77-89$K). This 
enables readout of large LArTPC detectors at the very low noise level of 
ENC~<~1000~e$^-$. Further, because the data is digitized near the wires, considerations of cable length to the warm electronics feedthrough are minimized, allowing CE and TPC designs to proceed mostly independently.. 
Additionally the CE includes warm interface electronics, which provide
local power control and real-time diagnostic readout at the cryostat, 
which was used to perform the validation and integration tests from
individual CE Boxes, to fully integrated APAs in the Cold Box, to the
completed TPC inside the cryostat.\\
120 CE Boxes were installed on 6 APAs by April 2018 for 15,360 
total TPC readout channels. These CE Boxes were selected for installation 
from a comprehensive QC testing effort. This QC effort also led to 
excellent CE performance during the installation and commissioning 
phases of detector operation, with 99.7\% of TPC channels active and 
92.8\% operating at very low levels of ENC. The performance of the TPC readout 
was monitored with the diagnostic readout during the detector activation 
phase leading up to physics data-taking in September 2018 and found 
only 4 channels which became inactive in the CE during this period. Currently
ProtoDUNE-SP is taking cosmic ray data with very stable performance from
the CE. 


\acknowledgments

The work described in this paper would not have been possible without a large number
of collaborating institutions from the DUNE Collaboration, who provided shift leaders
and shift operators for all of the quality control tests at BNL and who contributed to the
installation and commissioning work at CERN. In particular we would like to thank: 
Boston University, 
University of California at Davis,
CERN,
University of Cincinnati,
Chung-Ahn University, 
Colorado State University,
Comsewogue High School,
Fermi National Accelerator Laboratory,
University of Florida, 
University of Houston, 
Iowa State University,
Kansas State University,
Lousiana State University,
Michigan State University, 
Stony Brook University,
University of Texas (Arlington),
Tsinghua University,
and the US Department of Energy Science Undergraduates Laboratory Internships program. 
We acknowledge the support from DOE (Department of Energy)  USA.
We would also like to specifically thank Roberto Acciarri, Flavio Cavanna, Karol Henessey, 
Giovanna Lehmann, Regina Rameika, Filippo Resnati, and Christos Touramanis for their
leadership during the preparation, installation, and commissioning of the ProtoDUNE-SP 
detector. A special thanks to Bill Miller and the Ash-River installation group for the 
excellent job they did during the installation and commissioning and to the CERN Neutrino Platform which provided enormous technician and engineering support.
The work of J. Calcutt and K. Mahn was supported by the Office of Science, the Office of High Energy Physics, and of the U.S. Department of Energy award de-sc0015903.
\appendix



\bibliographystyle{JHEP}

\bibliography{protodune_coldelec_paper}{}

\end{document}